%% file: UIST2025-FlueBricks.tex
\documentclass[sigconf]{acmart}
\input{_includes}

\AtBeginDocument{%
  \providecommand\BibTeX{{
    \normalfont B\kern-0.5em{\scshape i\kern-0.25em b}\kern-0.8em\TeX}}}

\copyrightyear{2026}
\acmYear{2026}
\setcopyright{cc}
\setcctype{by}
\acmConference[CHI '26]{Proceedings of the 2026 CHI Conference on Human Factors in Computing Systems}{April 13--17, 2026}{Barcelona, Spain}
\acmBooktitle{Proceedings of the 2026 CHI Conference on Human Factors in Computing Systems (CHI '26), April 13--17, 2026, Barcelona, Spain}
\acmPrice{}
\acmDOI{10.1145/3772318.3790595}
\acmISBN{979-8-4007-2278-3/2026/04}

\def\note[#1#2#3]{#1\if b#2$\flat_#3$\else\if#2##$\sharp_#3$\else$_#2$\fi\fi}

\let\oldcite\cite
\renewcommand{\cite}[1]{\mbox{\oldcite{#1}}}

\begin{document}

\title[FlueBricks]{FlueBricks: A Construction Kit of Flute-like Instruments for Acoustic Reasoning}

\author{Bo-Yu Chen}
\affiliation{
  \institution{National Taiwan University}
  \city{Taipei}
  \country{Taiwan}
}
\author{Chiao-Wei Huang}
\affiliation{
  \position{Independent Researcher}
  \city{Taipei}
  \country{Taiwan}
}
\author{Lung-Pan Cheng}
\orcid{0000-0002-7712-8622}
\affiliation{
  \institution{National Taiwan University}
  \streetaddress{No. 1, Sec. 4, Roosevelt Rd.}
  \city{Taipei}
  \country{Taiwan}
  \postcode{10617}
}

\begin{abstract}
We present FlueBricks, a construction kit for \textit{acoustic reasoning} via building and customizing flute-like instruments. 
By assembling \textit{generator}, \textit{resonator}, and \textit{connector} modules that embody various aeroacoustic properties, users gain deeper understanding of how blowhole, tube length, and tone-hole placement alter onset, pitch, and timbre through hands-on experimentation. 
This forms a \textit{designer--player loop} of configuring and playing to form, test, and refine acoustic behaviors---\textit{acoustic reasoning}---shifting acoustic instruments from static artifacts to dynamic systems.
To understand how users engage with this system, we conducted an exploratory study with 12 participants ranging from novices to professional musicians. 
During their explorations, we observed participants fluently switching between designer and player roles, scaffolding designs from familiar instruments, forming and refining their acoustic understanding of length, tone holes, and generator geometry, reinterpreting modules beyond their intended functions, and using their creations for performative acts such as pedagogical showing and musical expression. 
These collectively demonstrated FlueBricks's potential as a pedagogical tool for embodied \textit{acoustic reasoning}. 
\end{abstract}

\ccsdesc[500]{Human-centered computing~User interface toolkits}

\keywords{Acoustic reasoning, instrument prototyping, acoustic construction kit, embodied interaction, tangible learning, flute}

\begin{teaserfigure}
  \includegraphics[width=\textwidth]{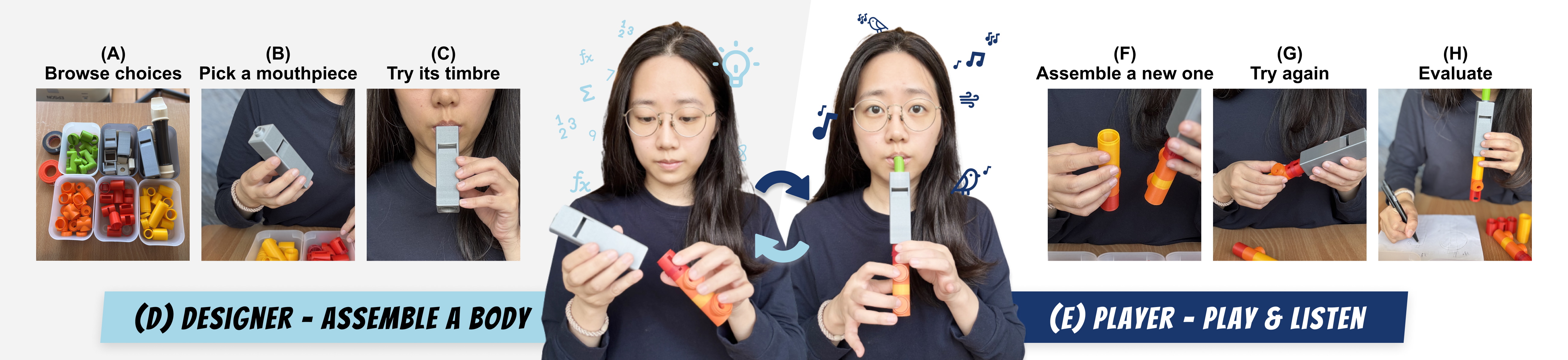}
  \caption{
    \textbf{FlueBricks and the designer--player loop.} FlueBricks is a construction kit of 3D-printed building blocks for building and customizing flute-like instruments. This figure illustrates the \textit{designer--player loop}: users iteratively switch between designing instruments and playing them to form, test, and refine acoustic behaviors. (A--C) Users browse tangible building blocks that embody aeroacoustic properties, select a mouthpiece, and breathe into it to hear the sound it generates. (D) As designers, they connect modules to assemble a resonator body. (E) As players, they immediately test the sound by playing. (F--H) Based on what they hear, users reconfigure tangible modules to test new hypotheses (F\&G) and play again to evaluate the sound effect (H). Through these iterations---hypothesizing, configuring, playing, and refining---users engage in \textit{acoustic reasoning}.
  }
  \Description{
    FlueBricks and the designer--player loop. FlueBricks is a construction kit of 3D-printed building blocks for building and customizing flute-like instruments. This figure illustrates the designer--player loop: users iteratively switch between designing instruments and playing them to form, test, and refine acoustic behaviors. (A--C) Users browse tangible building blocks that embody aeroacoustic properties, select a mouthpiece, and breathe into it to hear the sound it generates. (D) As designers, they connect modules to assemble a resonator body. (E) As players, they immediately test the sound by playing. (F--H) Based on what they hear, users reconfigure tangible modules to test new hypotheses (F and G) and play again to evaluate the sound effect (H). Through these iterations---hypothesizing, configuring, playing, and refining---users engage in acoustic reasoning.
  }
  \label{fig:figure1}
\end{teaserfigure}
\maketitle

\section{Introduction \& Related Work}
\label{sec:introduction}

Tangible prototyping toolkits~\cite{Hartmann2006,Greenberg2001} have long been one of the cores in technical Human-Computer Interaction (HCI), enabling non-experts to engage in hands-on learning experiences~\cite{Papert1980} and rapidly go through design iterations~\cite{Norman2013}. 
Many modalities, such as visual, haptic, and even olfactory, have been embodied as tangible toolkits~\cite{Dominiak2024,Feick2023,Lei2022}.

Auditory interaction, on the other hand, is inherently tangible---physical instruments date back to prehistoric times, with the Divje Babe flute among the earliest known examples~\cite{turk1995oldest}. Yet the relationship between instrument makers and players has long been bifurcated across history~\cite{Montagu2017,Rogers1977}.
The main reason is that instrument making requires complex acoustic tuning processes involving specialized craftsmanship~\cite{Terrien2013,Pal2006}, while acoustic performance has been characterized as a dynamic coupling where the body and instrument mutually shape the resulting sound~\cite{OModhrain2018}.
\textit{Acoustic reasoning}---an iterative process that requires both hands-on reconfiguration and play to form, test, and refine acoustic behaviors---has since fallen into oblivion in elementary musical pedagogy.

Digital music tools reshape this relationship. Synthesizers, samplers, and software-based instruments offer flexible sound manipulation, enabling users to design, modulate, and perform sound within a single system~\cite{Tahrolu2020}. 
This blurs traditional roles and has fueled the development of Digital Musical Instruments (DMIs) that emphasize modularity and live control~\cite{Kvifte2008,Ward2023} to achieve new musical effects.
Bringing multisensory, physical models with haptic devices into digital systems~\cite{Leonard2020} further
shows us the need to embed acoustic reasoning in new hybrid instruments.

With the advancement of 3D printing technology, the relationship comes even closer as more people are now able to make instruments in digital ways, speeding up the development of modular instruments.
For example, Modular Fiddle~\cite{OpenFabPDX2025} allows users to customize components of electronic violins to suit their personal style.
LeMo~\cite{Arbel2022} explores modularity in acoustic instruments as an educational assembly kit, decomposing solid-body instruments into \textit{generators}, \textit{resonators}, and \textit{radiators} for hands-on construction.

Flute-like instruments, historically simple yet culturally pervasive~\cite{Southcott2016Early,YamahaCorporationndSchool,YamahaCorporationndCreating}, have also attracted growing attention. 
Acoustic researchers have investigated how small geometric changes such as the flue, window, labium, and tuning slots systematically affect onset reliability, timbre, and pitch~\cite{Fletcher1998,Mercer1951,Douglas1965}.
In the meantime, more digital technology has been employed to design and make flute-like instruments.
For example, Printone~\cite{Umetani2016} provides sophisticated resonance modeling, and Acoustic Voxels~\cite{Li2016} optimizes resonator geometry via simulation.
3D printed flutes~\cite{Zoran2011,Kolomiets2020} are typically modeled in CAD software prior to fabrication, while Nicolas Bras' \textit{Modular Flute}~\cite{ModularFluteNicolas2025} offers preset mouthpiece and body sections for performance flexibility.

In this paper, we extend the concept of LeMo~\cite{Arbel2022} by bringing finer-grained aeroacoustic properties from acoustic research of flue-like instruments into an assembly kit for hands-on exploration. Specifically, we design \textit{acoustically meaningful} modules that expose more parameters than \textit{generators}, \textit{resonators}, and \textit{radiators} to support deeper acoustic reasoning. 
We present FlueBricks, a construction kit for acoustic learning via building and customizing flute-like instruments using 3D-printed building blocks. 
Figure~\ref{fig:figure1} shows how users work with FlueBricks.
FlueBricks offers several tangible building blocks that embody the aeroacoustic properties of flute-like instruments.
Users browse these modular components (Fig.~\ref{fig:figure1}A) and start by picking a mouthpiece for testing (Fig.~\ref{fig:figure1}B).
By breathing directly into the module, users hear the sound it generates (Fig.~\ref{fig:figure1}C). 
With its tangibility, it induces users' curiosity to start connecting with other modules, making them switch roles to instrument designers (Fig.~\ref{fig:figure1}D). 
As soon as users put on another module, they immediately switch back as players and test the sound again (Fig.~\ref{fig:figure1}E). 
This keeps users in the iteration of reconfiguring tangible modules (Fig.~\ref{fig:figure1}F\&G) and responsively evaluating the sound effect by playing their current prototypes (Fig.~\ref{fig:figure1}H). 

To understand how users with various levels of musical and woodwind expertise engage with FlueBricks, we conducted a 12-participant exploratory user study. Our results show that FlueBricks support acoustic reasoning: participants formed hypotheses about how structural changes would affect sound, configured modules to test those predictions, played to observe outcomes, and refined their understanding accordingly. These manifested in behaviors such as (1) instrument scaffolding (grounding designs in familiar instruments), (2) rule formation (articulating principles connecting geometry to sound), (3) reinterpretation (repurposing modules beyond intended functions), and (4) performative acts (consolidating exploration into pedagogical demonstrations and musical expression).

\subsection*{Contribution}

FlueBricks not only brings LeMo's~\cite{Arbel2022} concept of an assembly kit for hands-on education to flute-like instruments but also explores how computational acoustic design tools~\cite{Umetani2016,Li2016} can be rearticulated as modular, tangible toolkits~\cite{Dominiak2024,Feick2023,Lei2022}.

In contrast to existing 3D-printed modular instruments~\cite{OpenFabPDX2025, ModularFluteNicolas2025}, FlueBricks decomposes flute-like instruments into finer acoustical units informed by organ-pipe~\cite{Douglas1965} and flute acoustics~\cite{Fletcher1998}, exposing tube length, tone-hole size and position, and sound hole geometry as tangible parameters to support \textit{acoustic reasoning} in which users iteratively switch between instrument makers and players to refine acoustic behaviors.

More specifically, FlueBricks materializes acoustic parameters in reconfigurable parts, enabling users to explore airflow, resonance, and tone production through tangible manipulation rather than purely digital modeling~\cite{Wiberg2018}. This mirrors computational thinking---hypothesis, testing, revision---in craft-like interaction~\cite{Wiberg2018,Lindell2013}, and aligns with experience-centered perspectives that emphasize felt, exploratory engagement over abstract control~\cite{McCarthy2004}.
We see FlueBricks as an embodied way to explore future computational instrument designs, despite being physical artifacts that contain no computational unit.
We first describe the FlueBricks system and its usage configurations, then report findings from an exploratory user study, and conclude with a theoretical account of acoustic reasoning and design implications for constructive musical tools.

\begin{figure*}[!ht]
\centering
  \includegraphics[width=1\linewidth]{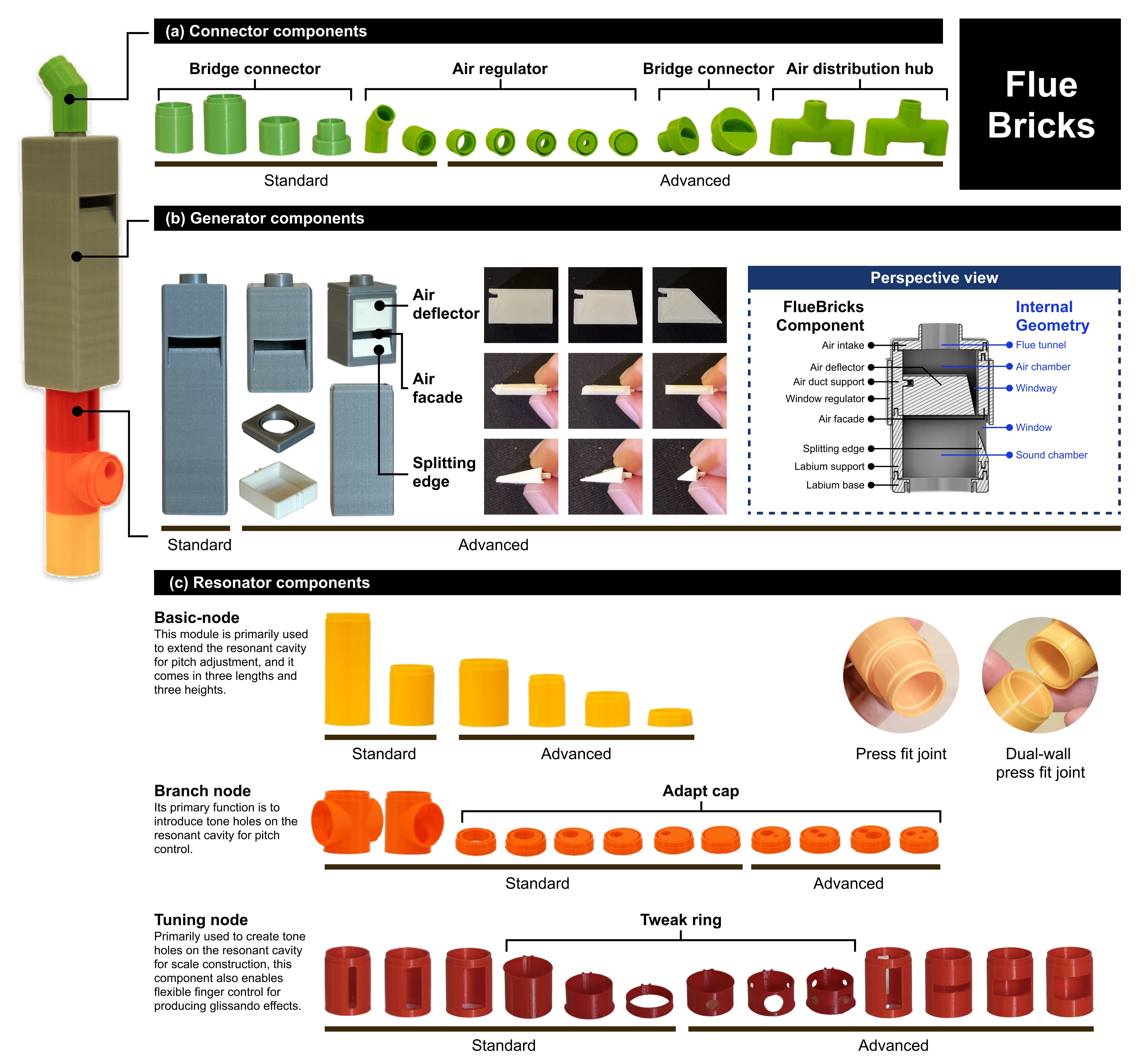}
  \caption
  {
    \textbf{Overview of the FlueBricks modular component system.}
     A complete flute constructed from FlueBricks modules, comprising three component families---\textbf{Generator}, \textbf{Resonator}, and \textbf{Connector}---each exposing parameter-level control for acoustic reasoning. The system offers two tiers: a \textit{standard set} of pre-assembled variants for getting started, and an \textit{advanced set} for fine-grained acoustic manipulation. A zoomed-in sketch illustrates the Generator's internal structure, highlighting its modularity and granularity.
  }
  \Description
  {
    Overview of the FlueBricks modular component system. A complete flute constructed from FlueBricks modules, comprising three component families---Generator, Resonator, and Connector---each exposing parameter-level control for acoustic reasoning. The system offers two tiers: a standard set of pre-assembled variants for getting started, and an advanced set for fine-grained acoustic manipulation. A zoomed-in sketch illustrates the Generator's internal structure, highlighting its modularity and granularity.
  }
  \label{fig:handbook}
\end{figure*}

\section{FlueBricks}
\label{sec:fluebricks}

We grounded the design of FlueBricks in Ledo et al.'s framework~\cite{Ledo18}, focusing on four goals: (G1) reducing authoring time and complexity, (G3) empowering new audiences, (G4) integrating with existing practices, and (G5) enabling replication and creative exploration. We excluded (G2) creating paths of least resistance to prioritize open-ended exploration over prescriptive guidance.
FlueBricks achieves these four goals by (1) enabling rapid iteration through plug-and-play tangible assembly with the levels of granularity, (2) providing standard modules for novices while maintaining advanced modules for experts, (3) resembling traditional flute-like instruments, and (4) enabling users to scaffold designs from known instruments while engaging in acoustic reasoning.

Figure~\ref{fig:handbook} shows an overview of FlueBricks, a modular toolkit that decomposes flute-making into interchangeable components designed for acoustic reasoning. It consists of three component families: \textit{generators}, which produce the initial tone; \textit{resonators}, which shape pitch and timbre; and \textit{connectors}, which link modules or hybridize with existing recorders. 
This aligns with the modular taxonomy proposed in LeMo~\cite{Arbel2022}, except that we omit radiators as flutes radiate through their body and holes, and add connectors for additional flexibility (G5).

For each component type presented below, acoustic design rationale and detailed specifications are available in the appendix (see Appendix~\ref{app:acoustic-rationale} for design rationale and Appendix~\ref{app:module-specifications} for module specifications).

\subsection{Generator Components}
\label{subsec:generator-components}

\begin{figure}[!htb]
\centering
  \includegraphics[width=1\columnwidth]{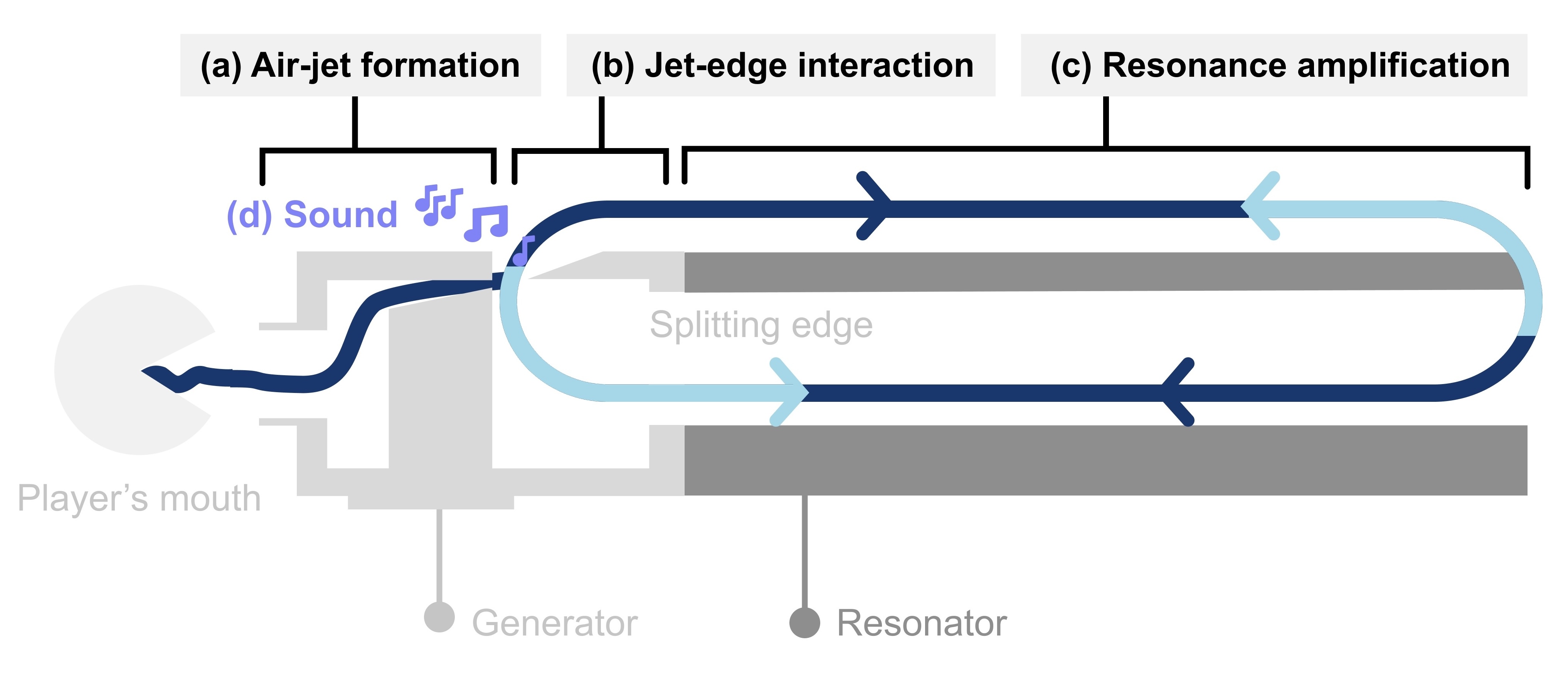}
  \caption{%
    \textbf{Three coupled phases of flute-like sound production.}
    (a) \textit{Air-jet formation}: blowing into the mouthpiece creates a focused jet of air. 
    (b) \textit{Jet-edge interaction}: the air jet encounters the labium, a sharp edge, and oscillates between the inside and outside of the pipe, producing an edge tone.
    (c) \textit{Resonance amplification}: The jet-labium interaction generates pressure oscillations that excite the resonator, which selectively reinforces certain frequencies through resonance to sustain the sound.
    These acoustic mechanisms inform FlueBricks's generator design: following the granularity principle, we decompose the generator into 8 acoustically meaningful submodules that map to these geometries (flue tunnel, air chamber, windway, window, splitting edge, sound chamber), enabling parameter-level control over tone production.
  }
  \Description{
    Three coupled phases of flute-like sound production.
    (a) Air-jet formation: blowing into the mouthpiece creates a focused jet of air.
    (b) Jet-edge interaction: the air jet encounters the labium, a sharp edge, and oscillates between the inside and outside of the pipe, producing an edge tone.
    (c) Resonance amplification: The jet-labium interaction generates pressure oscillations that excite the resonator, which selectively reinforces certain frequencies through resonance to sustain the sound.
    These acoustic mechanisms inform FlueBricks's generator design: following the granularity principle, we decompose the generator into 8 acoustically meaningful submodules that map to these geometries (flue tunnel, air chamber, windway, window, splitting edge, sound chamber), enabling parameter-level control over tone production.
  }
  \label{fig:flute_mechanism}
\end{figure}

Flute-like instruments generate sound by transforming breath into self-sustained oscillation through structural geometry~\cite{Fletcher1998}. As shown in Fig.~\ref{fig:flute_mechanism}, the generator condenses breath into an air jet that interacts with a splitting edge to initiate oscillation, which is then amplified by the resonator. Unlike reed or brass instruments, flutes achieve oscillation purely through geometry. 

Flutes have long been studied in acoustics because small geometric variations can dramatically alter tonal qualities such as onset, pitch, and timbre~\cite{Fletcher1998}. From this literature, we identified 6 internal geometries most critical to tone: the \textit{flue tunnel}, which sets airflow volume and loudness~\cite{Steenbrugge2010}; the \textit{air chamber}, which shapes onset reliability and timbre~\cite{Mercer1951}; the \textit{windway}, where dimensions influence jet stability and tonal color~\cite{Nolle1979}; the \textit{window}, whose height and profile tune pitch and brightness~\cite{Nolle1979, Hruska2021}; the \textit{splitting edge}, which governs oscillation onset~\cite{Mercer1951}; and the \textit{sound chamber}, which defines resonance and loudness~\cite{Halfpenny1956}.
To achieve (G1), we trade off between the number of modules and the granularity of acoustic levers and decompose the generator into 8 submodules that map to the following acoustic geometries: \textit{Air intake}, \textit{Air duct support}, \textit{Air deflector}, \textit{Air facade}, \textit{Window regulator}, \textit{Splitting edge}, \textit{Labium support}, and \textit{Labium base}. 

Based on this structure, FlueBricks offers two tiers: a \textit{standard set} of pre-assembled all-in-one variants for immediate playability, and an \textit{advanced set} of modular submodules for fine-grained control. This achieves (G3) by lowering barriers for novices while enabling experts to manipulate acoustic levers directly.

\subsection{Resonator Components}
\label{subsec:resonator-components}

\begin{figure}[!htb]
\centering
  \includegraphics[width=1\columnwidth]{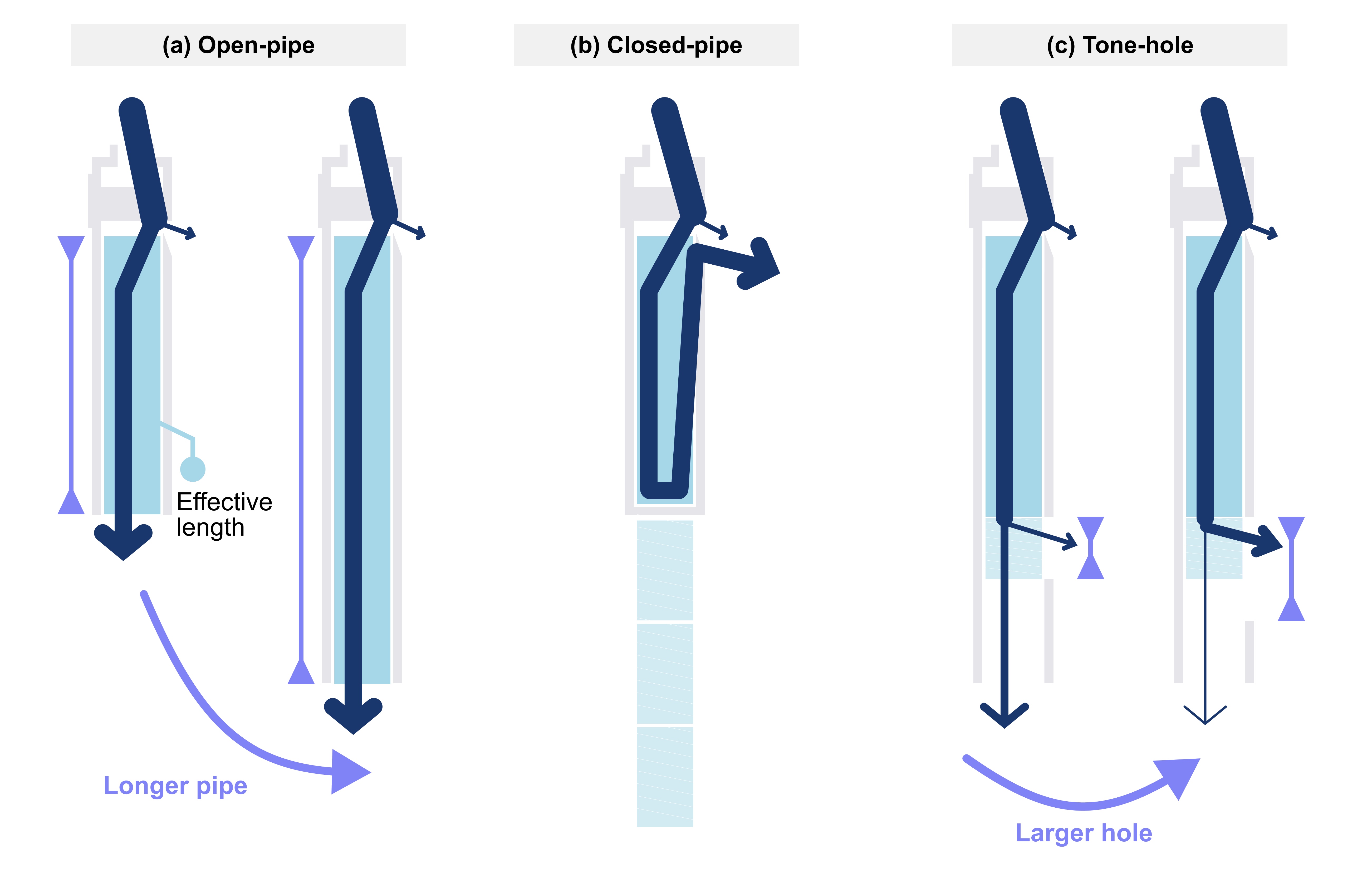}
  \caption{%
    \textbf{How changes in pipe geometry affect pitch.}
    This figure compares three types of resonator configurations. 
    Left: an open pipe creates sound by supporting vibrations along its full length, with both ends open. 
    Middle: a closed pipe supports a different vibration pattern, resulting in a longer effective length and a lower pitch compared to an open pipe of the same size. 
    Right: a tone hole acts like a shortcut: if large enough, it defines a new endpoint for vibration, effectively shortening the pipe and raising the pitch. 
    The cyan region shows the effective length---the part of the pipe that actually shapes the sound.
  }
  \Description{
    How changes in pipe geometry affect pitch.
    This figure compares three types of resonator configurations.
    Left: an open pipe creates sound by supporting vibrations along its full length, with both ends open.
    Middle: a closed pipe supports a different vibration pattern, resulting in a longer effective length and a lower pitch compared to an open pipe of the same size.
    Right: a tone hole acts like a shortcut: if large enough, it defines a new endpoint for vibration, effectively shortening the pipe and raising the pitch.
    The cyan region shows the effective length---the part of the pipe that actually shapes the sound.
  }
  \label{fig:resonator_mechanism}
\end{figure}

Like many wind instruments, flutes control pitch by adjusting the effective length of the resonating pipe~\cite{Fletcher1998}. As illustrated in Fig.~\ref{fig:resonator_mechanism}, an open pipe vibrates along its full length, while a closed pipe extends the effective air column, lowering the fundamental. Tone holes work differently: their position sets where the air column effectively ends, while their size influences how strongly they shorten the pipe~\cite{Fletcher1998}. Hole size and coverage influence pitch and tone color. Advanced techniques such as glissando enable continuous pitch variation through gradual modification of effective length~\cite{Nesterova_2020}.

By modularizing tube length, hole placement, and cavity form, FlueBricks turns resonance into an intuitive design space, enabling recorder-like replication (G4) as well as creative exploration (G5).
To achieve these goals, FlueBricks introduces a set of \textit{resonator nodes} that make length and opening variations tangible. 
A single node can be extended or shortened to set the air-column length, or outfitted with a tone hole whose position, size, and degree of coverage define pitch. 
By stacking nodes, users construct fingering systems, rearrange scales, and combine holes of different sizes to achieve bends and microtonal variation. 

To achieve (G3), we provide both discrete (\textit{basic nodes}) and continuous (\textit{telescope}) options to offer multiple resolution levels. \textit{Basic nodes} provide stable, fixed-length starting points that scaffold discrete length-to-pitch exploration for novices. \textit{Telescope nodes} enable continuous fine-tuning but require two-handed manipulation. 
As shown in Fig.~\ref{fig:handbook}, we design five resonator modules, each mapped to an acoustic lever: \textit{Basic node}, \textit{Branch node}, \textit{Adapt cap}, \textit{Tuning node}, and \textit{Tweak ring}.
These modules also form a system for finer-grained pitch resolution, enabling custom pitch systems that bridge recorder-style fingering (G4) with exploratory scales, microtonality, and timbral variation (G5).

\subsection{Connector Components}
\label{subsec:connector-components}

Unlike the \textit{Generator} and \textit{Resonator}, which directly produce and shape sound, the \textit{Connector} acts as an infrastructural backbone---supporting airflow routing, branching structures, and hybrid configurations. 
Connector modules allow a single breath to be divided across multiple chambers, enabling multi-voice instruments and sustained drones~\cite{Collinson1975}, or linked into existing instruments for hybrid setups. 
As illustrated in Fig.~\ref{fig:handbook}, the connector family includes three modules---\textit{air distribution hub}, \textit{air regulator}, and \textit{bridge connector}---which together embody FlueBricks's emphasis on (G4). These connectors enable acoustic reasoning by allowing rapid reconfiguration for hypothesis testing.
These connectors extend FlueBricks beyond single-pipe instruments, aligning with established woodwind infrastructures (G4) while introducing ergonomic improvements (G1) and new airflow routing possibilities (G5).

\section{FlueBricks Usage Configurations}
\label{sec:fluebricks-usage-configurations}

This section illustrates how FlueBricks might be used through its modular design. The following configurations show intended usage patterns for generator, resonator, and connector components, establishing the baseline of designed interactions that participants then extended, reinterpreted, and creatively adapted in the user study (Section~\ref{sec:results}).

\begin{figure}[!htb]
\centering
  \includegraphics[width=1\columnwidth]{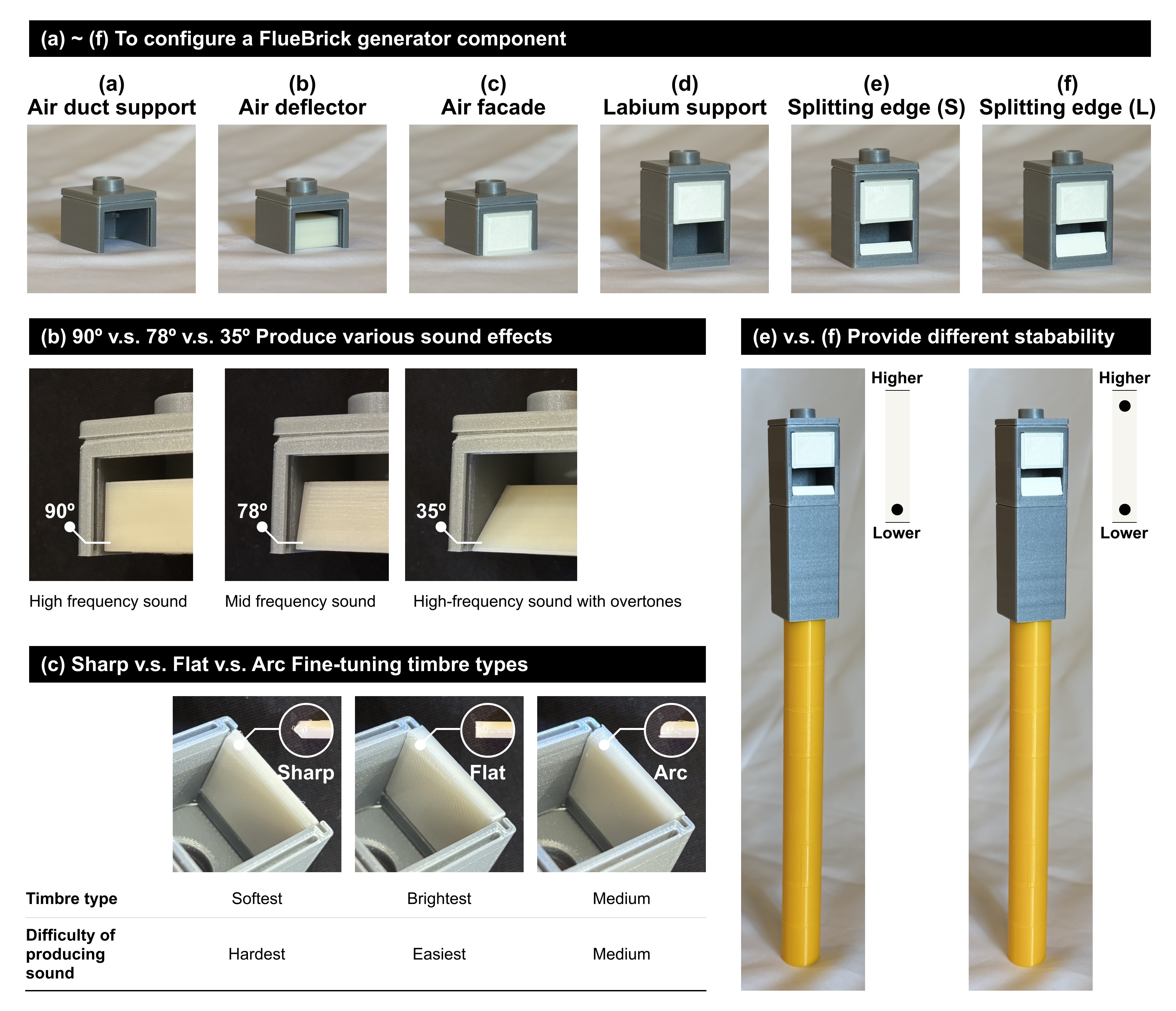} 
  \caption
  {%
    \textbf{Generator fine-grained control configuration.} FlueBricks enables airflow and tone production control through parameter-level manipulation. (a--f) The generator is assembled from eight modular parts, each controlling specific acoustic parameters. (b) Varying air deflector angles (90°, 78°, 35°) reshapes airflow, shifting tone from focused high frequencies to overtone-rich textures. (c) Changing the \textit{air facade} (sharp, flat, arc) affects timbre brightness and ease of onset. (e--f) A shorter \textit{splitting edge} raises the overblowing threshold, improving pitch stability in lower registers.
  }
  \Description
  {
    Generator fine-grained control configuration. FlueBricks enables airflow and tone production control through parameter-level manipulation. (a--f) The generator is assembled from eight modular parts, each controlling specific acoustic parameters. (b) Varying air deflector angles (90 degrees, 78 degrees, 35 degrees) reshapes airflow, shifting tone from focused high frequencies to overtone-rich textures. (c) Changing the air facade (sharp, flat, arc) affects timbre brightness and ease of onset. (e--f) A shorter splitting edge raises the overblowing threshold, improving pitch stability in lower registers.
  }
  \label{fig:walkthrough_6}
\end{figure}

 \textbf{Generator fine-grained control configuration:} Building on the \textit{advanced set} of modular submodules introduced in Section~\ref{subsec:generator-components}, users can manipulate the eight generator submodules (\textit{Air deflector}, \textit{Air facade}, \textit{Splitting edge}, etc.) to control airflow and tone production. As shown in Fig.~\ref{fig:walkthrough_6}, varying these units enables parameter-level control over timbre, onset, and pitch stability. This demonstrates (G5) by exposing fine-grained acoustic parameters that users can swap and test iteratively, starting from a standard generator and progressively refining tone characteristics through module substitution.

\begin{figure}[!htb]
\centering
  \includegraphics[width=1\columnwidth]{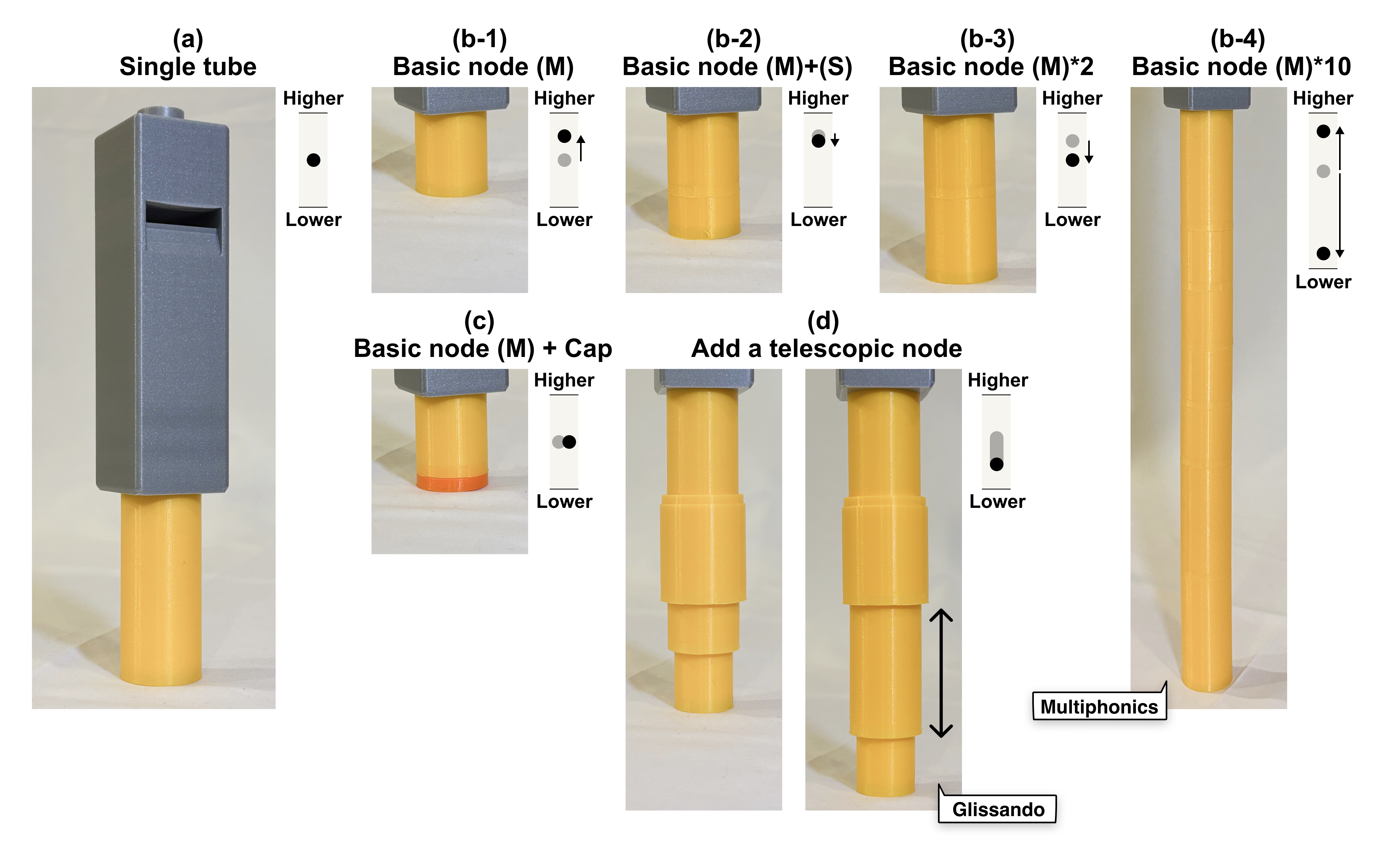} 
  \caption
  {
    \textbf{Resonator length control configuration.} FlueBricks enables pitch control through modular length manipulation. (a) \textit{Basic nodes} of different lengths provide discrete pitch steps. (b-1 to b-4) Stacking nodes extends the air column, progressively lowering pitch and deepening timbre. (c) Attaching an \textit{adapt cap} shifts to closed-pipe mode for further pitch lowering. (d) \textit{Telescoping nodes} enable continuous pitch bending for glissando effects.
  }
  \Description
  {
    Resonator length control configuration. FlueBricks enables pitch control through modular length manipulation. (a) Basic nodes of different lengths provide discrete pitch steps. (b-1 to b-4) Stacking nodes extends the air column, progressively lowering pitch and deepening timbre. (c) Attaching an adapt cap shifts to closed-pipe mode for further pitch lowering. (d) Telescoping nodes enable continuous pitch bending for glissando effects.
  }
  \label{fig:walkthrough_1}
\end{figure}

 \textbf{Resonator length control configuration:} Following the discrete and continuous options introduced in Section~\ref{subsec:resonator-components}, users control pitch through the \textit{Basic node} (discrete length steps) and telescoping nodes (continuous adjustment). As shown in Fig.~\ref{fig:walkthrough_1}, stacking nodes extends the air column, while attaching an \textit{Adapt cap} shifts to closed-pipe mode. This configuration supports both G4 (replication of recorder-like instruments through discrete nodes) and G5 (creative exploration through telescoping and closed-pipe configurations), enabling users to scaffold from familiar pitch relationships to novel tunings.

\begin{figure}[!htb]
\centering
  \includegraphics[width=1\columnwidth]{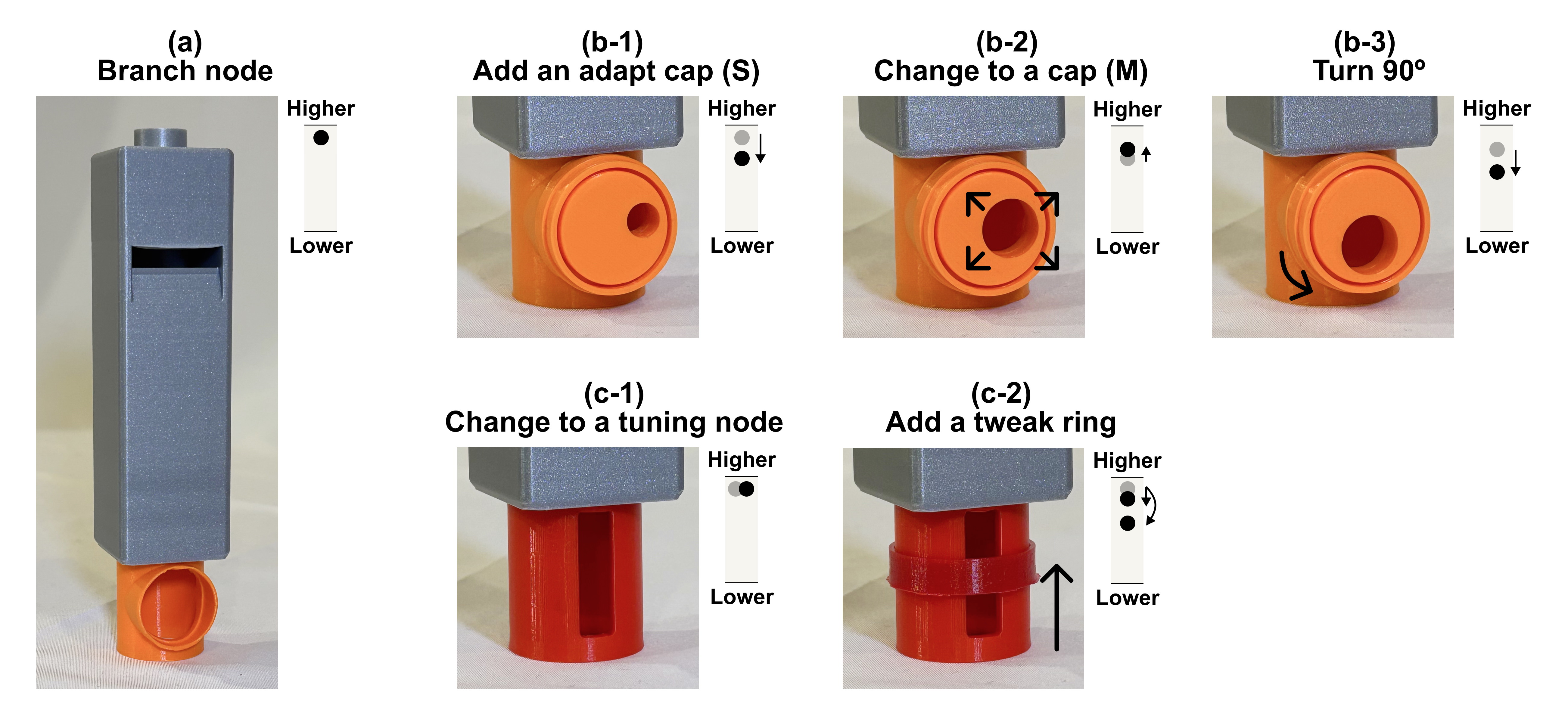} 
  \caption
  {%
    \textbf{Tone-hole manipulation configuration.} FlueBricks enables pitch and timbre control through hole size, position, and continuous adjustment. (a) A fully-opened \textit{branch node} creates a high pitch by shortening the effective resonator length. (b-1 to b-3) \textit{Adapt caps} adjust pitch by varying tone-hole size across caps and changing direction through rotation. (c-1) A \textit{tuning node} with a vertical slot enables broader tone shaping. (c-2) A sliding \textit{tweak ring} supports real-time pitch modulation for glissando effects.
  }
  \Description
  {%
    Tone-hole manipulation configuration. FlueBricks enables pitch and timbre control through hole size, position, and continuous adjustment. (a) A fully-opened branch node creates a high pitch by shortening the effective resonator length. (b-1 to b-3) Adapt caps adjust pitch by varying tone-hole size across caps and changing direction through rotation. (c-1) A tuning node with a vertical slot enables broader tone shaping. (c-2) A sliding tweak ring supports real-time pitch modulation for glissando effects.
  }
  \label{fig:walkthrough_2}
\end{figure}

 \textbf{Tone-hole manipulation configuration:} Building on the resonator modules introduced in Section~\ref{subsec:resonator-components}, users control pitch and timbre through the \textit{Branch node}, \textit{Adapt cap}, \textit{Tuning node}, and \textit{Tweak ring}. As shown in Fig.~\ref{fig:walkthrough_2}, these modules enable control over hole size, position, and continuous adjustment. Users might start by opening a branch node to raise pitch, then rotate an adapt cap to reposition the aperture, and finally slide a tweak ring for real-time glissando effects, demonstrating how multiple resolution levels (G5) support both discrete and continuous pitch exploration.

\begin{figure}[!htb]
\centering
  \includegraphics[width=1\columnwidth]{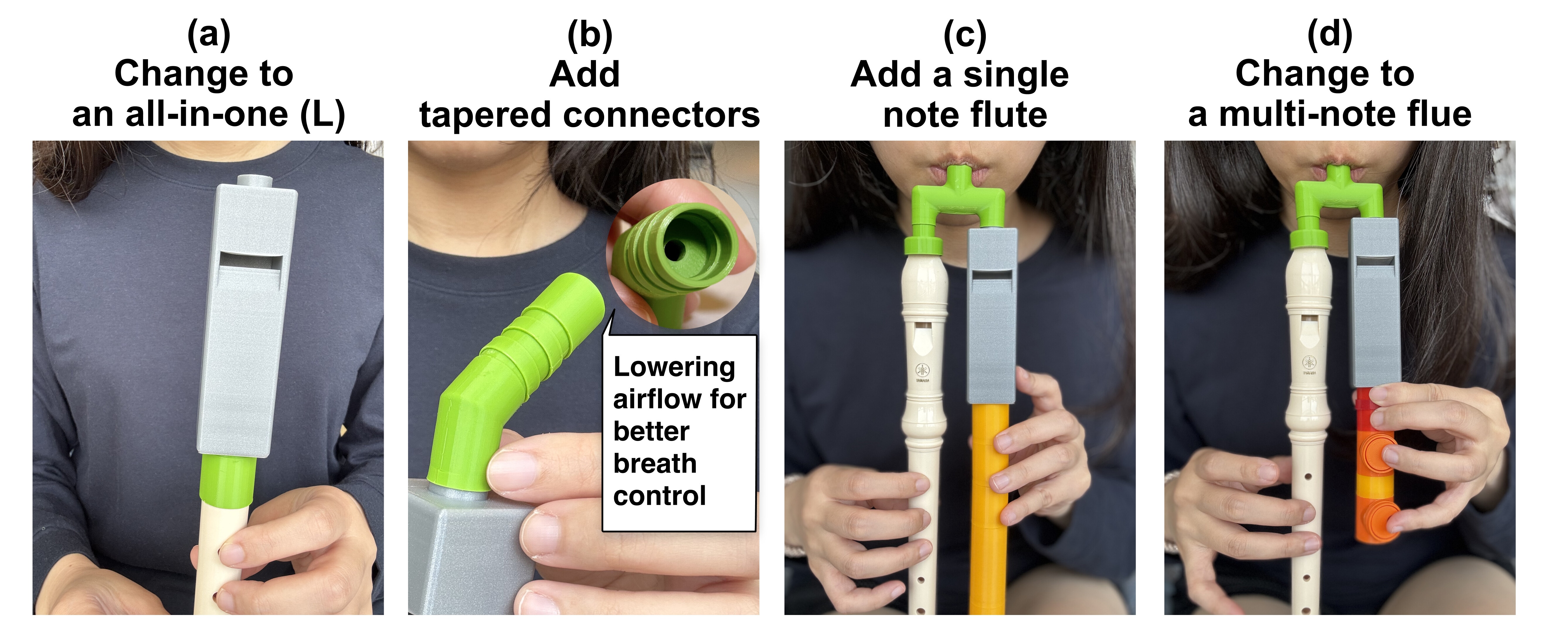} 
  \caption
  {%
    \textbf{Connector configurations and instrument hybridization.} (a) An \textit{all-in-one (L)} generator module is attached to create a distinct timbre. (b) \textit{Tapered connectors} are added to provide ergonomic angling and restrict airflow for better breath control. (c) A bifurcated \textit{air hub} integrates a standard recorder with a \textit{single-note flute} for drone accompaniment. (d) The setup is upgraded to a \textit{multi-note flute}, enabling simultaneous melody lines and harmonic complexity.
    }
  \Description
  {%
    Connector configurations and instrument hybridization. (a) An all-in-one (L) generator module is attached to create a distinct timbre. (b) Tapered connectors are added to provide ergonomic angling and restrict airflow for better breath control. (c) A bifurcated air hub integrates a standard recorder with a single-note flute for drone accompaniment. (d) The setup is upgraded to a multi-note flute, enabling simultaneous melody lines and harmonic complexity.
  }
  \label{fig:walkthrough_5_6}
\end{figure}

\textbf{Connector configuration:} Using the connector modules introduced in Section~\ref{subsec:connector-components} (\textit{air distribution hub}, \textit{air regulator}, and \textit{bridge connector}), FlueBricks enables two complementary approaches. The \textit{recorder hybridization configuration} (Fig.~\ref{fig:walkthrough_5_6}, left) demonstrates integration with existing instruments: bridge connectors ensure compatibility, while swapping generators and adding air regulators personalize timbre and ergonomics. The \textit{multi-voice design configuration} (Fig.~\ref{fig:walkthrough_5_6}, right) enables harmony through airflow routing: an air distribution hub splits airflow from a single breath source, allowing simultaneous drone and melody lines. Together, these configurations demonstrate G4 (integration with existing practices) and G5 (creative exploration) through modular connector configurations.
Section~\ref{sec:results} further shows how participants extended these configurations through instrument scaffolding, rule formation, and reinterpretation.

\section{Exploratory User Study Setup}
\label{sec:study-setup}
We conducted an exploratory user study to investigate user engagement with FlueBricks as both designers and players. Following our design rationale of excluding (G2), we prioritized exploration over prescriptive guidance to foster the \textit{designer--player loop}. Minimal guidance was provided to observe how participants with varying backgrounds naturally interpret and navigate between construction and performance roles. All sessions were conducted individually to enable focused observation of each participant's reasoning process.

\begin{figure*}[!t]
\centering
  \includegraphics[width=1\linewidth]{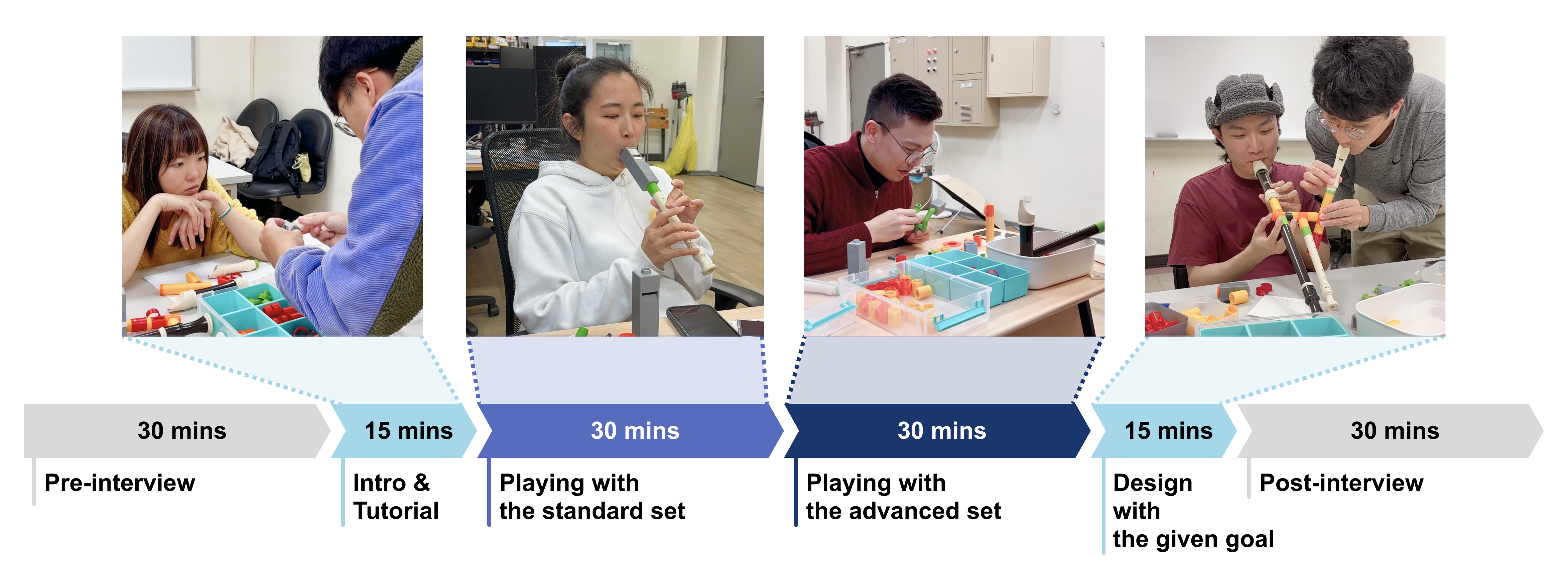}
  \caption
  {
    \textbf{Overview of the user study procedure.} The study consisted of six phases, alternating between interviews, open-ended exploration, and goal-directed design. The 90-minute hands-on exploration period enabled participants to engage in iterative hypothesize--configure--evaluate--conclude cycles, which we analyzed as evidence of acoustic reasoning (Section~\ref{sec:results}). While time blocks were planned, the schedule was flexibly adjusted based on each participant's engagement level and curiosity.
  }
  \Description
  {
    Overview of the user study procedure. The study consisted of six phases, alternating between interviews, open-ended exploration, and goal-directed design. The 90-minute hands-on exploration period enabled participants to engage in iterative hypothesize--configure--evaluate--conclude cycles, which we analyzed as evidence of acoustic reasoning. While time blocks were planned, the schedule was flexibly adjusted based on each participant's engagement level and curiosity.
  }
  \label{fig:study_setup}
\end{figure*}

\subsection{Participants}
\label{subsec:participants}

We recruited 12 participants, 7 males and 5 females, aged 20 to 51 (\textit{M}=31.9, \textit{SD}=7.5), evenly distributed across four distinct user profiles to ensure a range of musical and woodwind expertise: 

\begin{itemize}[leftmargin=*] 
\item\textbf{Novice} = no formal musical training or woodwind experience. 
\item\textbf{Amateur} = informal musical training with some experience playing woodwind instruments.
\item\textbf{Pro Woodwind} = formally trained in woodwind instruments.
\item\textbf{Pro Musician} = formally trained music composer and producer. 
\end{itemize}

Each expertise group consisted of three participants. Complete demographic information is detailed in Table~\ref{tab:participant_background}. Participants were compensated with 750 TWD for completing the 2-hour session.

\subsection{Procedure}
\label{subsec:procedure}

Each study session followed a six-phase process lasting approximately two hours (Fig.~\ref{fig:study_setup}). The researcher observed and responded to technical inquiries but did not guide design decisions.

The session began with a 30-minute pre-interview on musical background and building experience, followed by a 15-minute introduction covering only physical module connection. Participants then engaged in 30 minutes of unguided exploration with the standard set using a think-aloud protocol, followed by 30 minutes of advanced exploration with the full toolkit. A 15-minute design task required creating a ``classroom flute for elementary students.'' The session concluded with a 30-minute post-interview capturing reflections on strategies and discoveries. While the 90-minute hands-on exploration period was planned, the actual duration was flexibly adjusted based on each participant's engagement and curiosity. Sessions were video recorded and transcribed for thematic analysis.

\subsection{Metrics and Coding Scheme}
\label{subsec:metrics-coding-scheme}

We developed a structured coding scheme to assess \textbf{module-level acoustic use} and \textbf{participant-level perceptual and emotional responses} from think-aloud data, behavioral observation, and post-session interviews.

\paragraph{Module Use Categories}
Each participant-module interaction was classified according to its highest level of sound-related engagement:
\begin{itemize}
  \item \textbf{Non-Acoustic}: Structural or aesthetic use, with no acoustic intent.
  \item \textbf{Exploratory}: Sound-focused testing or verbal speculation, even if not included in the final design.
  \item \textbf{Intentional}: Deliberate acoustic use, indicated by integration into an instrument or verbal explanation.
\end{itemize}

This categorization is visualized in Fig.~\ref{fig:result_module_usage}.

\paragraph{Perceptual Attention and Affective Summary}
Four perceptual dimensions---Pitch Variation, Timbre Variation, Onset Sensitivity, and Sense of Control---were coded on a 3-point scale based on reference frequency: \textbf{(1)} low (0--2 times), \textbf{(2)} moderate (3--4), \textbf{(3)} high (5+). Each participant was assigned an \textbf{affective label} (Positive, Neutral, or Negative) based on the dominant emotional tone in their post-session interview, summarized in Fig.~\ref{fig:result_perception} right.

\section{Exploratory User Study Results}
\label{sec:results}

We analyzed participants' usage behaviors, perception patterns, and reflective interpretations through observational data and post-interview transcripts.

\subsection{Module Usage Patterns}
\label{subsec:module-usage-patterns}

\begin{figure}[!htb]
\centering
  \includegraphics[width=1\columnwidth]{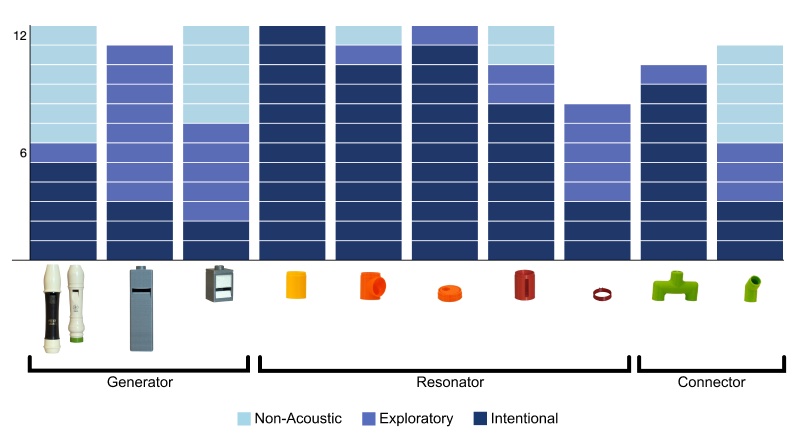} 
  \caption
  {
      \textbf{Module-level acoustic use across participants.} Each row represents a participant and each column a module. Colors indicate the highest level of sound-related use: Non-Acoustic (no acoustic intent), Exploratory (acoustic testing), or Intentional (sound-related function).
  }
  \Description
  {%
      Module-level acoustic use across participants. Each row represents a participant and each column a module. Colors indicate the highest level of sound-related use: Non-Acoustic (no acoustic intent), Exploratory (acoustic testing), or Intentional (sound-related function).
  }\label{fig:result_module_usage}
\end{figure}

Participants engaged extensively with the modular system, though the depth and nature of interaction varied across module categories (Fig.~\ref{fig:result_module_usage}, left).

\textit{Generators} were universally recognized as sound-producing components.
All participants interacted with at least one generator module, though the ways they approached and applied them varied considerably. Traditional generators (recorder-style mouthpieces) were intentionally used by 5 participants---all with wind instrument backgrounds---who emphasized their reliability and familiarity. As P2 explained, \textit{``I’m more familiar with it---it feels more reliable.''}
The \textit{all-in-one} generator was explored by 11 participants, but only 3 integrated it into final designs. Most used it primarily for experimenting with onset sensitivity and timbral variation, without a clearly defined acoustic objective.
The \textit{modular} generator saw the most evenly distributed usage: all 12 participants tried it, 5 explored internal variations, and 2 used it intentionally in their designs. Its modularity encouraged iterative play, though unstable airflow often posed challenges. Several participants began with the \textit{modular} generator but switched to the \textit{all-in-one} or traditional variant for more predictable tone production. Many noted difficulty comparing acoustic differences, as each reassembly reset the listening context. As P4 remarked, \textit{``Every time I change it, I forget how the last one sounded.''} Some suggested providing another \textit{modular} generator for side-by-side comparison. Others found the assembly effort discouraging, especially when acoustic differences were minimal. Several noted that without a clear sense of improvement, it became harder to justify continued adjustments.

\textit{Resonators} were the most widely adopted module category, appearing in nearly all final designs. The \textit{basic node} was intentionally used by all 12 participants, typically serving as the primary resonating body. \textit{Branch nodes} and \textit{adapt caps} were each intentionally used by 11 participants, often combined to form tone-hole-like structures for pitch modulation. These modules offered strong physical affordances; as P8 noted, \textit{``This looks like it gives us a tone hole.''} Many participants used multiple branch nodes simultaneously, exploring non-linear paths for tone control. This playful use of spatial configuration not only reflects a departure from traditional recorder-style designs, but also enabled novel acoustic experiments.
The \textit{tuning node} was intentionally used by 8 participants, with 2 others exploring it without integrating it into their final designs. While some appreciated its length-adjusting role, others found it difficult to operate---often because they overlooked the \textit{tweak ring}, which was designed to stabilize it during play. As P8 noted, \textit{``Ah! I have to use both hands to cover everything---then it kind of feels not worth it.''} Only 3 participants (P1, P4, and P12) used the \textit{tuning node} and \textit{tweak ring} in combination with clear acoustic intent. This group spanned different levels of musical background, suggesting that hands-on experimentation was especially important when a module’s physical affordance was ambiguous or lacked cues inherited from flute-like instruments.
The \textit{tweak ring} was explored by 8 out of 12 participants but intentionally used by only 3. Lacking obvious form-function cues, it was frequently misinterpreted as a connector, stabilizer, or decorative piece. As P6 joked, \textit{``Is this just for decoration?''} Its intended role---enabling smooth length adjustment when paired with the \textit{tuning node}---was not immediately apparent.
Three participants (P5, P7, and P11) independently reinterpreted the \textit{tweak ring} as a sliding element for pitch control. They inserted it into slit-style tone holes and experimented with real-time motion to modulate pitch, later discovering that this use aligned with its originally intended function of adjusting hole size. None of them included this behavior in their final design.

\textit{Connectors} in this section focus on modules with potential acoustic function. The \textit{bridge connector} is excluded, as it was intended purely for structural linkage. While some participants explored unconventional acoustic uses for it, these reinterpretations are discussed in Section~\ref{subsec:reinterpretation} on emergent behaviors.
The \textit{air distribution hub} was explored by 11 and intentionally used by 10 participants---surpassing even the tuning node. Unlike most other modules, the hub has no direct counterpart in traditional recorder design. Yet its radial form prompted playful experimentation with airflow routing and tonal divergence. Some participants used it to produce multiple simultaneous tones by directing air into separate generators. As P4 observed, \textit{``It’s like two flutes playing at once.''} In one instance, when the participant needed to test a complex configuration, the interviewer briefly collaborated by blowing into a shared generator routed through the hub, creating a combined airflow. 
Several participants described the hub as visually interesting or worth experimenting with, even before understanding its function. A similar response was observed with the \textit{bridge connector}, which drew early attention due to its asymmetry and was often tested playfully.
In many cases, these reinterpretations were carried forward into final designs---often to reinterpret the hub as a resonator. As P2 noted, \textit{``It has nice pitch control,''} after experimenting with enclosing different outlets and redirecting internal air paths.
The \textit{air regulator}, by contrast, followed a more restrained usage pattern. Most participants treated it as a breath adapter and stopped experimenting once it stabilized airflow. Only four participants used it with explicit acoustic intent---two as unconventional resonators and one (P5) as a muffler for softening tone. Others overlooked its varied hole-size configurations entirely. As P1 reflected, \textit{``I thought it just made blowing easier---I didn’t realize it changed the sound.''}

Overall, modules with strong physical affordances or familiar forms---such as recorder-style generators and tone-hole-like branches---were readily adopted. In contrast, less transparent or unconventional components, like the \textit{tweak ring} and \textit{air distribution hub}, prompted wider exploration and reinterpretation.

\begin{figure}[!htb]
\centering
  \includegraphics[width=1\columnwidth]{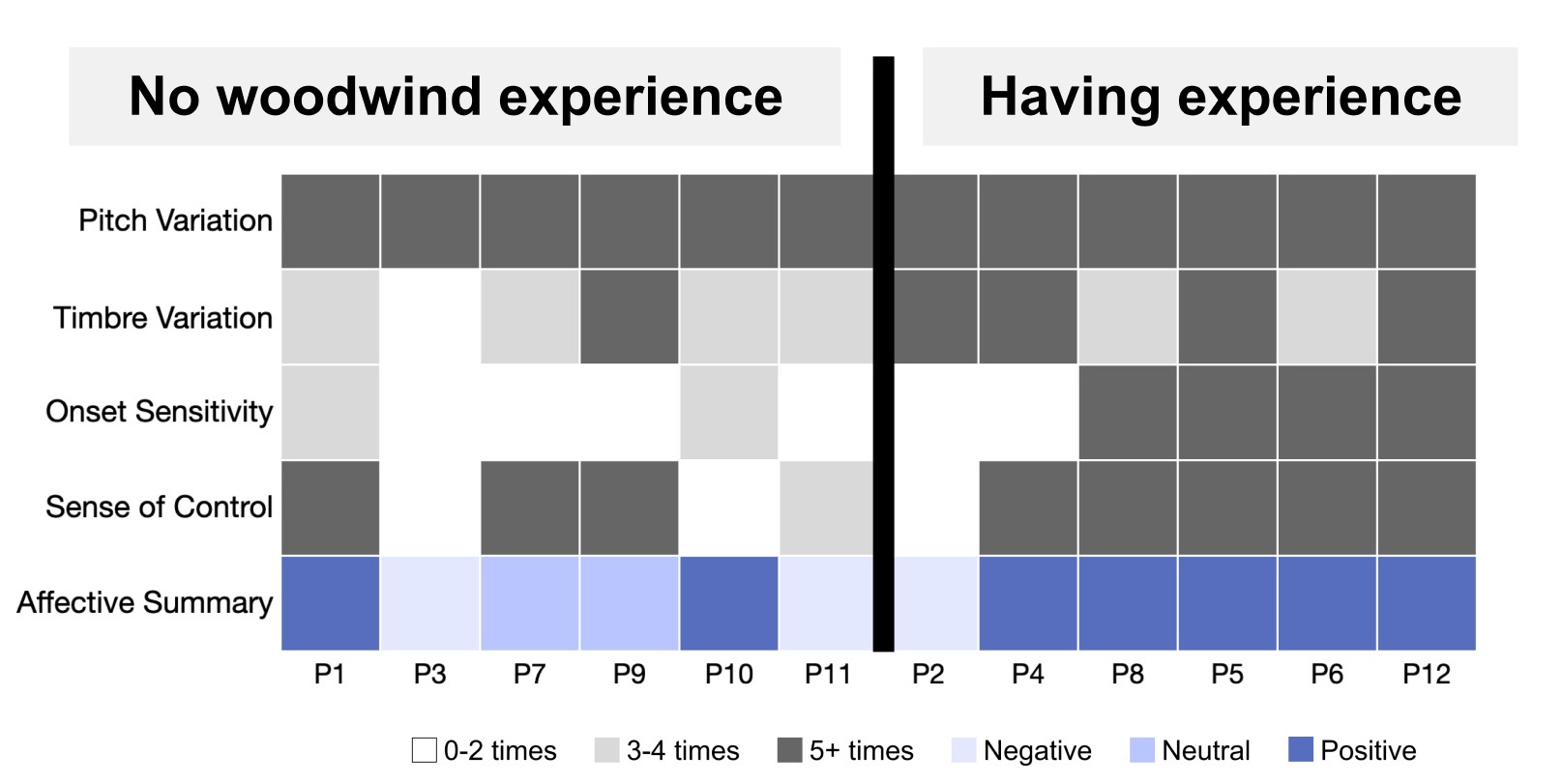} 
  \caption
  {%
    \textbf{Participant-level perceptual and emotional responses.}
    Each column represents a participant, grouped by musical or woodwind background.
    Rows indicate \textit{Pitch Variation}, \textit{Timbre Variation}, \textit{Onset Sensitivity}, and \textit{Sense of Control}; the bottom row shows an affective summary (Positive, Neutral, Negative).
    Color intensity encodes frequency ranges (0--2, 3--4, 5+ times).
  }
  \Description
  {%
    Participant-level perceptual and emotional responses.
    Each column represents a participant, grouped by musical or woodwind background.
    Rows indicate Pitch Variation, Timbre Variation, Onset Sensitivity, and Sense of Control; the bottom row shows an affective summary (Positive, Neutral, Negative).
    Color intensity encodes frequency ranges (0--2, 3--4, 5+ times).
  }
  \label{fig:result_perception}
\end{figure}

\subsection{Perceptual Sensitivity and Exploratory Experience}
\label{subsec:perceptual-sensitivity}

Participants exhibited high perceptual engagement across sound-related dimensions, though sensitivity varied by expertise (Fig.~\ref{fig:result_perception}). 
\textbf{Pitch variation} was universally perceived across all participants regardless of background, serving as the most accessible entry point for acoustic reasoning. As P1 noted, \textit{``I managed to make something similar to a soprano recorder---something that can produce a sense of scale or pitch variation.''} However, Fig.~\ref{fig:result_perception} reveals a distinct expertise gap in higher-order acoustic dimensions.

\textbf{Timbre} and \textbf{onset sensitivity} appear densely clustered among participants with woodwind experience. Timbre variation was often prompted by structural contrasts---such as open vs.\ closed pipes, or the difference between all-in-one and recorder generators---and by fine-tuning through modular mouthpiece adjustments. Onset sensitivity, while least observed overall, was consistently reported by all three pro woodwind participants (P6, P7, P12). P8, though not a specialist, also exhibited high onset awareness, attributing this to his prior experience with overblowing on bassoon and saxophone.

This visual distinction extends to the \textbf{Sense of Control}. The heatmap (Fig.~\ref{fig:result_perception}) illustrates that control is not determined by general musical expertise, but specifically by woodwind experience.
Most woodwind amateurs and professionals---except P2---demonstrated high control and correspondingly positive or neutral experiences. However, P10 and P11, despite their musical expertise, expressed lower control. P11, in particular, described the experience as frustrating and unpredictable: \textit{``I didn't know what was working or why\ldots{} it felt random''}, and was the only musician with a negative summary. Conversely, P10, while unable to explain several acoustic effects, still enjoyed the ambiguity: \textit{``It didn't behave like I thought, but that made it fun to explore.''}

\begin{figure}[!htb]
\centering
  \includegraphics[width=1\columnwidth]{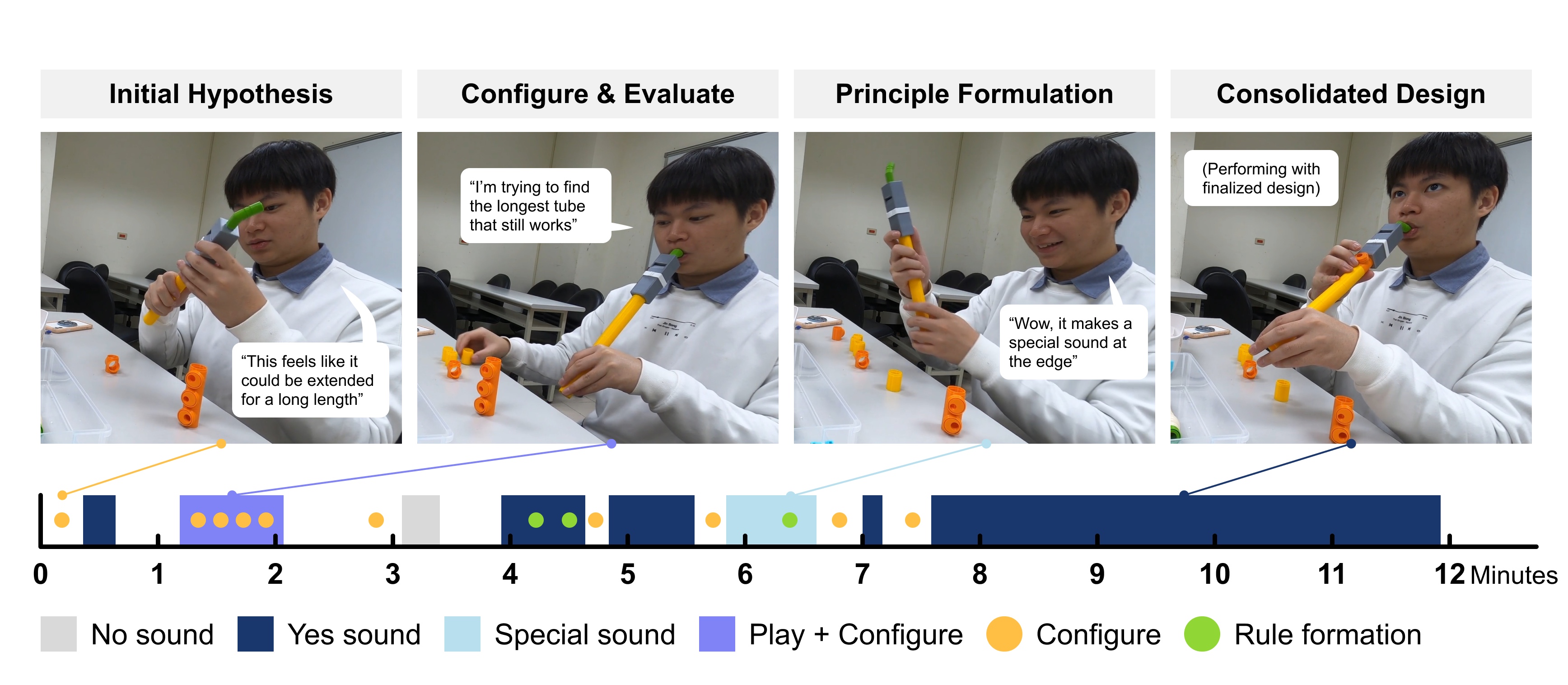} 
  \caption
  {%
    \textbf{The designer--player loop in action.} P12's (Pro Woodwind) session illustrates the \textit{designer--player loop}: rapid alternation between configuring and playing, with each sonic outcome informing the next adjustment.
  }
  \Description
  {%
    The designer--player loop in action. P12's (Pro Woodwind) session illustrates the designer--player loop: rapid alternation between configuring and playing, with each sonic outcome informing the next adjustment.
  }
  \label{fig:result_designer_player}
\end{figure}

\subsection{From Designer--Player Loop to Acoustic Reasoning}
\label{subsec:designer--player-loop}

We observed that participants' engagement was not only about switching roles between designer and player but also about forming and testing hypotheses---a process we frame as \textit{acoustic reasoning}, the iterative process of forming, testing, and refining hypotheses about acoustic behavior through hands-on configuration and play. Participants fluidly alternated between configuring, playing, and reconfiguring, with each sonic outcome serving as evidence to refine their understanding. P12's session (Fig.~\ref{fig:result_designer_player}) illustrates this process.

P12 (Pro Woodwind) began by probing the limits of resonator length, repeatedly extending the basic node until tonal stability was lost. This resulted in a distinctive \textit{play-while-configuring} behavior, where blowing and adjusting happened simultaneously. As he explained, \textit{``I'm trying to find the longest tube that still works---if I go too far, the sound drops out.''} When the configuration approached failure, he shifted strategies, using overblowing to generate layered textures near the resonance threshold, a technique sometimes described as \textit{multiphonics}. Surprised by the outcome, he remarked, \textit{``Wow, it makes a special kind of sound right at the edge---somewhere between two tones.''}

P12 then experimented with pitch modulation by moving a branch node and adapt cap. Initially, placing the hole near the foot yielded minimal effect, but relocating it closer to the mouthpiece enabled stronger bending. Reflecting on this discovery, he observed, \textit{``Putting it here doesn't do much\ldots{} but up here, I can bend the pitch more clearly. So it's not just the hole size---it's where you put it that matters.''} The session culminated in a custom nine-note setup that combined airflow articulation, lateral tone control, and finger placement. Summarizing the outcome, he concluded, \textit{``This is my best setting---I can control it with both breath and finger holes.''}

P12's session illustrates how iterative configuring and playing can evolve into an \textit{acoustic reasoning cycle}: beginning with a hypothesis, testing it through configuration and sound, evaluating the outcome, and consolidating it into a lightweight rule---a cycle we summarize as \textit{hypothesize, configure, evaluate, conclude}.

\begin{figure*}[!t]
\centering
  \includegraphics[width=\textwidth]{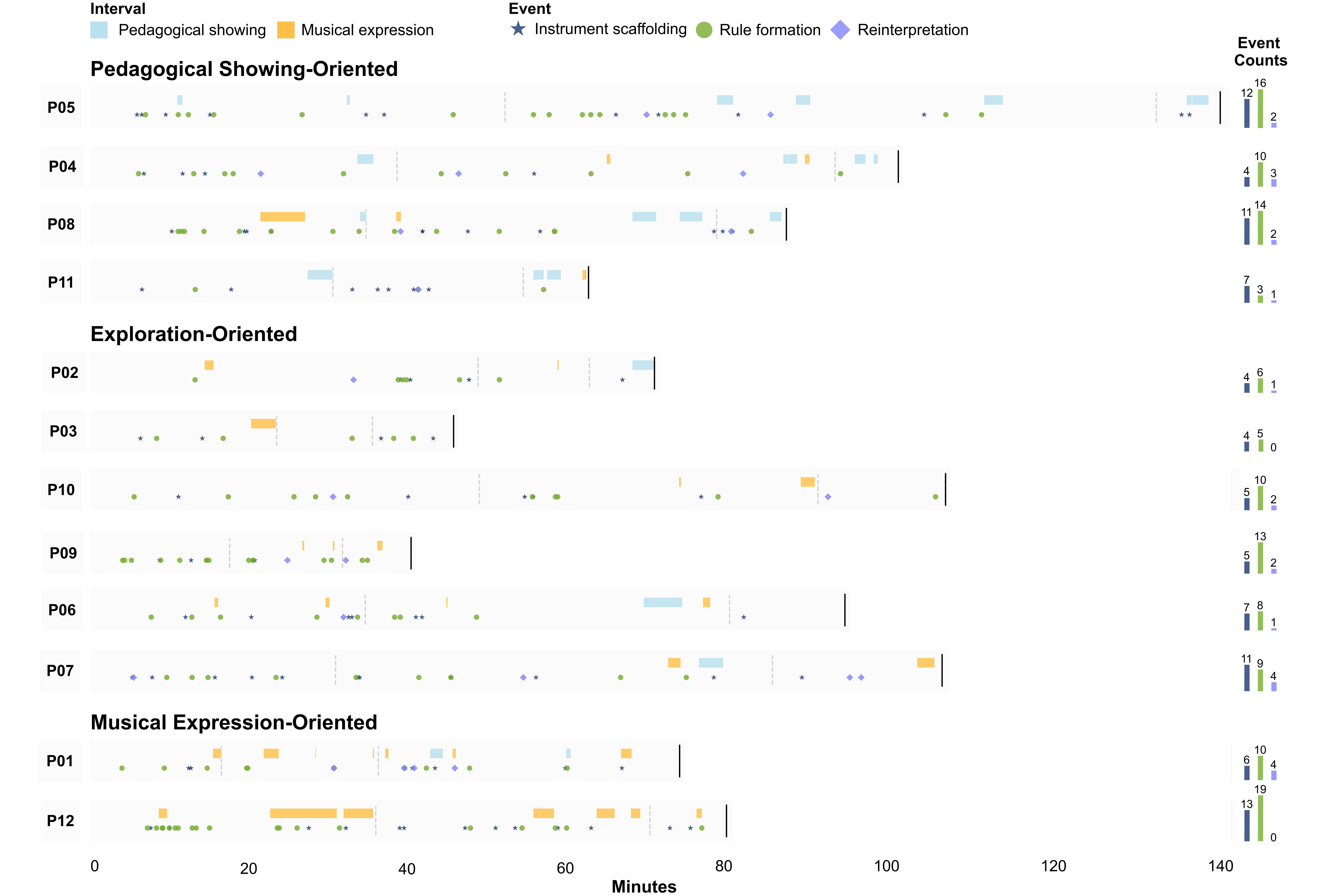}
  \caption
  {%
    \textbf{Exploration timelines of all 12 participants.} Each row shows one participant's session across three phases (dashed lines); point markers denote reasoning events---\textit{instrument scaffolding} (blue stars), \textit{rule formation} (green circles), and \textit{reinterpretation} (purple diamonds); horizontal bars denote duration-based \textit{performative acts}; right-side charts summarize per-behavior frequency.
  }
  \Description
  {%
    Exploration timelines of all 12 participants. Each row shows one participant's session across three phases (dashed lines); point markers denote reasoning events---instrument scaffolding (blue stars), rule formation (green circles), and reinterpretation (purple diamonds); horizontal bars denote duration-based performative acts; right-side charts summarize per-behavior frequency.
  }
  \label{fig:timeline}
\end{figure*}

Zooming out from this single case, Fig.~\ref{fig:timeline} presents a consolidated timeline of all 12 participants' exploration sessions (ending at the completion of the scenario task), revealing diverse exploration styles. Most participants spent the most time in advanced exploration, followed by basic and then scenario phases; P1 was an exception, becoming more engaged with a clear scenario goal. We categorized these journeys based on performative acts into three profiles: \textit{pedagogical showing-oriented}, \textit{musical expression-oriented}, and \textit{exploration-oriented}. Regardless of individual variations, three emergent behaviors appeared across all groups: \textit{instrument scaffolding}, \textit{rule formation}, and \textit{reinterpretation}.

\subsubsection{Instrument Scaffolding}
\label{subsec:instrument-scaffolding}

We define \textit{instrument scaffolding} as the way participants drew on familiar instruments to ground their reasoning and design decisions in FlueBricks, specifically involving acoustic grounding---linking module configurations to the sounds, resonances, and playing behaviors of instruments they already knew.

Patterns in Fig.~\ref{fig:timeline} show that instrument scaffolding varied substantially across participants. P5, P8, and P12 exhibited the highest frequency, drawing heavily on recorder and flute knowledge---P5 and P12 as professional woodwind musicians applying embodied techniques, P8 as an experienced amateur transferring techniques from harmonica and bassoon. In contrast, P9, P10, and P11 showed notably low scaffolding usage, prioritizing discovery of \textit{``new''} or \textit{``unfamiliar''} sounds over anchoring in known instruments, likely because working with flexible digital instruments had cultivated a preference for discovering unfamiliar sounds over reproducing known ones.

Within an \textit{acoustic reasoning cycle}, instrument scaffolding appeared in four forms. In \textit{intention setting}, participants framed goals in reference to known instruments (P1: \textit{``I want to make something like a recorder''}). In \textit{evaluation}, participants compared outcomes to familiar acoustic behaviors (P8: \textit{``This feels closer to a Dizi''}). In \textit{design guidance}, participants borrowed acoustic features from existing instruments, such as P11 attaching a basic node to a recorder body to extend its range, or P6 noting that recorder-style modules \textit{``make the tone more stable.''} In \textit{expressive scaffolding}, participants transferred playing techniques---P12 experimented with recorder-style overblowing, P7 applied familiar fingerings, and P4 attempted a transverse flute embouchure.

Across phases, scaffolding shifted from \textit{descriptive} to \textit{explanatory} to \textit{creative intent}, often coinciding with \textit{rule formation}. In summary, instrument scaffolding followed the \textit{acoustic reasoning cycle}---from framing goals through familiar instruments (\textit{hypothesize}), through selecting configurations based on known acoustic behaviors (\textit{configure}) and comparing outcomes to expected sounds (\textit{evaluate}), to consolidating designs grounded in instrument knowledge (\textit{conclude}).

\subsubsection{Rule formation}
\label{subsec:rule-formation}

We define \textit{rule formation} as the process by which participants articulated generalizable acoustic principles from their exploration, regardless of whether those principles were scientifically accurate or later revised. Our focus is on the behavior of rule articulation itself---the act of inferring and negotiating causal relationships between FlueBricks configurations and their acoustic outcomes. Participants often modified, rejected, or replaced earlier rules as new evidence emerged, and some rules were only partially correct; however, accuracy is not the focus of our analysis. Rule formation occurred when participants discovered new relationships, tested or updated prior beliefs, explained why certain configurations failed, or recognized how one component influenced another.

Patterns in Fig.~\ref{fig:timeline} show that rule formation was the most prevalent behavior across the study, suggesting that FlueBricks naturally supported iterative sense-making. P5, P8, and P12 demonstrated sustained rule articulation, reflecting continuous attempts to relate structural manipulations to acoustic effects; in contrast, P2, P3, and P11 show sparse, fragmented rule formation. Phase-wise, the basic exploration stage contains the densest early clusters, as participants frequently reasoned about resonator length, edge geometry, and other foundational relationships. Later phases show more distributed patterns, reflecting the reuse of earlier discoveries rather than frequent new findings. Sustained articulation correlated with higher perceived control, while fragmented patterns aligned with lower reported control. P10 presents an exception---his frequent rule formation coincided with low perceived control but positive affect, suggesting he treated it as open-ended inquiry rather than principle-building for design.

Examples of rule formation appeared throughout the study. P5 hypothesized, \textit{``If I make the tube longer, maybe the sound will drop,''} before adjusting the module to test the prediction. P2 compared tubes systematically (\textit{``I want to test them one by one''}), demonstrating methodical evaluation. P6 observed that adjusting the splitting edge improved onset stability (\textit{``Ah, now it speaks more easily''}), linking geometry to playability. Participants varied in their reasoning styles: some built knowledge through incremental micro-cycles, while others formed rules only after broader exploratory jumps. As Fig.~\ref{fig:timeline} shows, scaffolding often seeded rule formation hypotheses, and reinterpretation emerged when breaking an earlier rule enabled new uses of a module. In summary, rule formation followed the \textit{acoustic reasoning cycle}---from predicting acoustic relationships (\textit{hypothesize}), through systematic testing (\textit{configure}) and causal interpretation (\textit{evaluate}), to consolidating reusable principles (\textit{conclude}).

\subsubsection{Reinterpretation}
\label{subsec:reinterpretation}

We define \textit{reinterpretation} as repurposing FlueBricks modules beyond their intended acoustic roles, often sparked by ambiguous shapes or tactile feedback that invite uses breaking from expected functions.

Patterns in Fig.~\ref{fig:timeline} show that reinterpretation was the least frequent of the three reasoning behaviors. P1 and P7 exhibited the highest frequency, each producing four events; P4 also showed notable engagement with three events. All three were non-professional musicians whose exploratory confidence and lack of constraining prior models enabled them to probe unconventional uses. At the opposite extreme, P3 (novice) and P12 (professional) both showed zero events---the former constrained by uncertainty, the latter by established models---suggesting reinterpretation emerges most readily in a middle zone between inexperience and expertise.

\begin{figure}[!htb]
\centering
  \includegraphics[width=1\columnwidth]{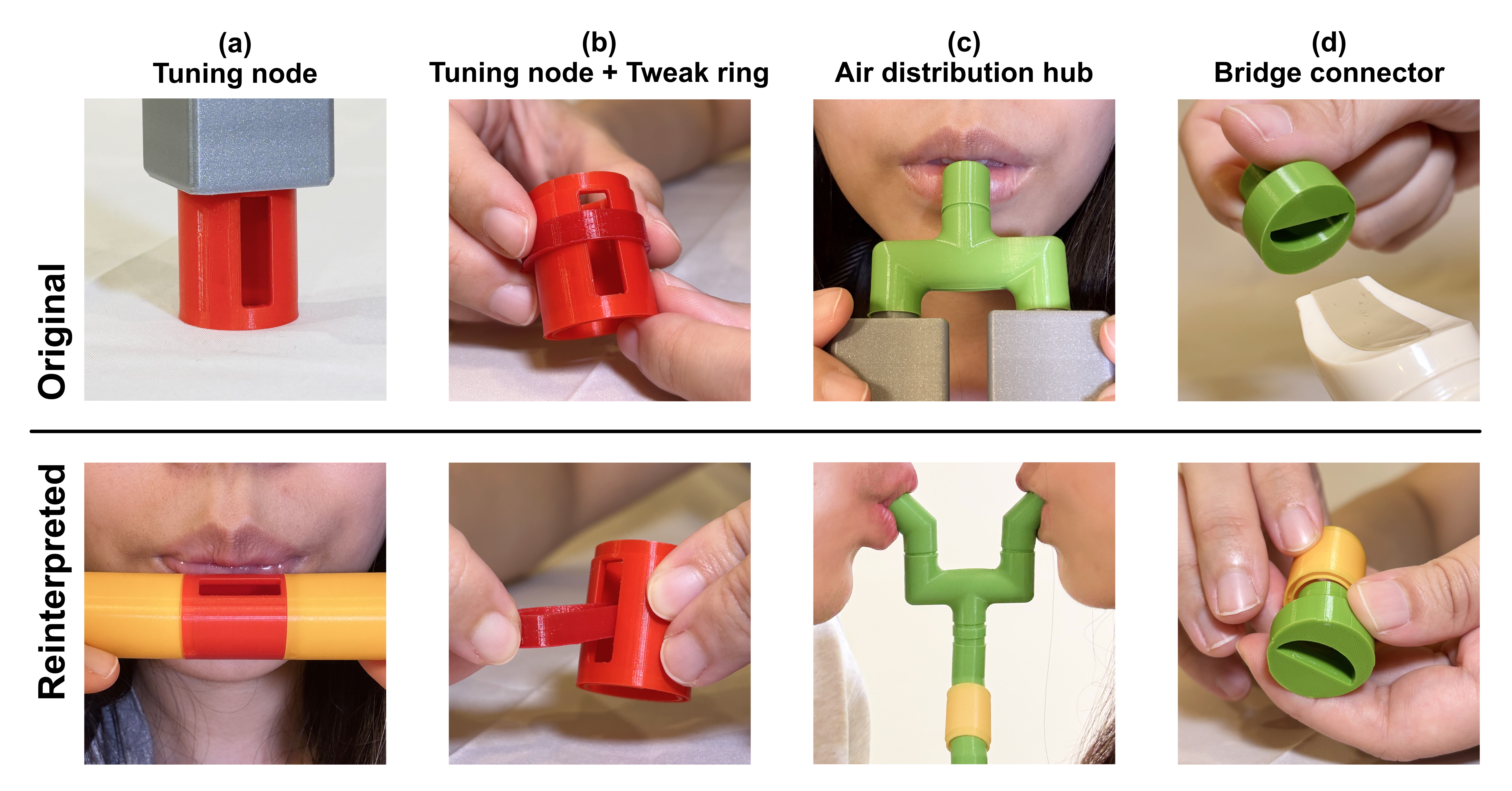} 
  \caption
  {%
    \textbf{User reinterpretations of FlueBricks components.} Despite their original functions---(a) \textit{tuning node}, (b) \textit{tweak ring}, (c) \textit{air hub}, and (d) \textit{bridge connector}---participants repurposed them creatively, revealing unexpected uses and modular flexibility.
  }
  \Description
  {%
    User reinterpretations of FlueBricks components. Despite their original functions---(a) tuning node, (b) tweak ring, (c) air hub, and (d) bridge connector---participants repurposed them creatively, revealing unexpected uses and modular flexibility.
  }
  \label{fig:result_reinterpreted}
\end{figure}

Examples of reinterpretation illustrate how ambiguous affordances enabled new acoustic functions (Fig.~\ref{fig:result_reinterpreted}). Some emerged through structural analogy: the \textit{tuning node}, designed as a pitch-adjustable resonator, was reimagined by P6 (a professional \textit{dizi} player) as a standalone generator---\textit{``Just this piece alone can form the structure of a dizi.''} Others arose through embodied play: the \textit{tweak ring}, intended for fine adjustment of resonator length, was used by P5, P7, and P11 as a glissando controller inserted into tone holes, enabling continuous pitch modulation. P8 reinterpreted the \textit{bridge connector}---originally a rigid joiner---as a free-floating pitch slider: \textit{``This piece is so responsive\ldots{} it lets me play all kinds of pitches.''} P2 and P9 repurposed the \textit{air distribution hub} and \textit{air regulator} as resonators by redirecting airflow or enclosing openings. P1 configured the \textit{air distribution hub} for dual-player use, transforming it into a collaborative generator.

In summary, reinterpretation followed the \textit{acoustic reasoning cycle}---from questioning alternative purposes (\textit{hypothesize}), through novel arrangements (\textit{configure}) and assessing acoustic effects (\textit{evaluate}), to establishing reusable functions (\textit{conclude}). Though less common than scaffolding or rule formation, reinterpretation revealed how users leveraged ambiguity to explore beyond intended design.

\subsection{Performative Acts}
\label{subsec:performative-act}

Beyond individual reasoning cycles, participants often consolidated their exploration into \textit{performative acts}---using their instruments as media for expression or teaching. These acts were not merely endpoints but natural extensions of \textit{acoustic reasoning}. We distinguish two forms: \textit{pedagogical showing}, where participants used their creations as teaching tools to demonstrate acoustic principles; and \textit{musical expression}, where participants configured modules into playable systems to perform melodies or timbral effects.

\begin{figure*}[t]
\centering
  \includegraphics[width=\textwidth]{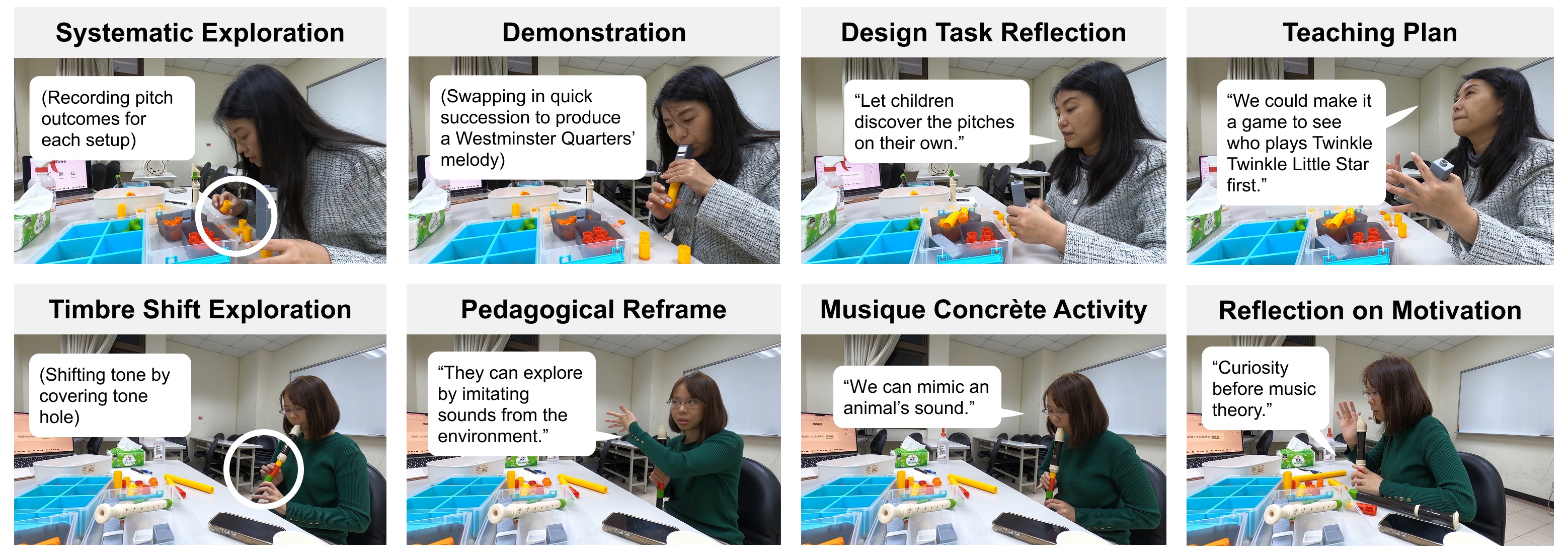} 
  \caption
  {%
      \textbf{Pedagogical showing as performative acts.} Top row: P5 (Pro Woodwind) used systematic pitch recording, melody demonstration, and a classroom game proposal to build a structured teaching strategy. Bottom row: P11 (Pro Musician) explored timbre manipulation and proposed open-ended discovery activities.
  }
  \Description
  {%
    Pedagogical showing as performative acts. Top row: P5 (Pro Woodwind) used systematic pitch recording, melody demonstration, and a classroom game proposal to build a structured teaching strategy. Bottom row: P11 (Pro Musician) explored timbre manipulation and proposed open-ended discovery activities.
  }
  \label{fig:pedagogical}
\end{figure*}

\subsubsection{Pedagogical showing}

Participants consolidated their exploration into teaching-oriented framings. In one example (Fig.~\ref{fig:pedagogical}, top row), P5 (Pro Woodwind) began with systematic exploration, carefully recording pitch outcomes for each configuration. Building on this, she gave a musical demonstration by swapping resonators in quick succession to reproduce the Westminster Quarters melody. Her reflection shifted toward education, noting that \textit{``children should discover the pitches on their own.''} Finally, she proposed a concrete teaching plan: framing instrument design as a playful challenge by asking students to compete over who could assemble and play Twinkle Twinkle Little Star first.

A different example (Fig.~\ref{fig:pedagogical}, bottom row) reveals a more imaginative approach to pedagogy. Here, P11 (Pro Musician) focused on timbre manipulation, experimenting with tone-hole coverage to shift sound quality. She reframed this exploration pedagogically, suggesting that learners could \textit{``imitate sounds from the environment.''} Extending this idea, she proposed musique concr\`{e}te activities such as mimicking animal sounds with the toolkit. She concluded by articulating a broader teaching philosophy---\textit{``curiosity before music theory''}---emphasizing the value of open-ended discovery over rigid instruction.

While some participants framed their discoveries as structured tasks and games, others leaned toward open-ended reframing and metaphor---both drawing on their acoustic explorations to propose concrete teaching scenarios.

\subsubsection{Musical expression}

\begin{figure*}[t]
\centering
  \includegraphics[width=\textwidth]{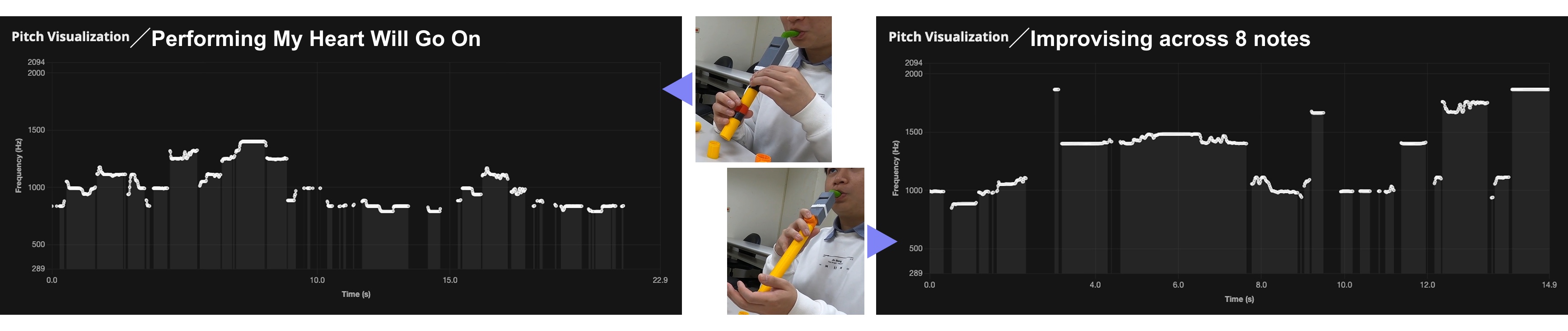} 
  \caption
  {%
    \textbf{Musical expression as performative acts.} Left: P12 (Pro Woodwind) performing the melody of \textit{My Heart Will Go On} with a custom multi-note setup. Right: P12 improvising freely across eight distinct pitches. Pitch contours were extracted using \textsc{SwiftF0}~\cite{nieradzik2025swiftf0}.
  }
  \Description
  {%
    Musical expression as performative acts. Left: P12 (Pro Woodwind) performing the melody of My Heart Will Go On with a custom multi-note setup. Right: P12 improvising freely across eight distinct pitches. Pitch contours were extracted using SwiftF0.
  }
  \label{fig:musical}
\end{figure*}

P12 (Pro Woodwind) demonstrated the most advanced examples of musical expression within FlueBricks. He first explored the \textit{branch node} configuration (Fig.~\ref{fig:result_designer_player}, right)---the outcome of an earlier \textit{designer--player loop}---achieving an 8-note improvisation (Fig.~\ref{fig:musical}, right) by coordinating breath velocity with simultaneous manipulation of the top branch hole and the resonator's end hole.

However, he subsequently switched to a tuning-node design, finding it easier to manipulate pitch. Comparing the two, he noted: \textit{``This is better than the other one\ldots{} because we can only control how much portion of tone cover to control the pitch.''} Applying a rule formulated during his earlier exploration---that geometry closer to the excitation source exerts stronger pitch influence---he deliberately arranged the tuning node immediately after the generator, followed by a basic node and a second tuning node. He consolidated this structure with tape to ensure airtight stability. After exploring the range, he realized the potential to play \textit{My Heart Will Go On}, transposing the melody down when the original key exceeded the instrument's range. Explaining this adaptive process, he noted: \textit{``Because we learn a lot of different instruments\ldots{} I am familiar with finding a fingering by my own based on my own sense.''} Fig.~\ref{fig:musical} (left) visualizes the stable melodic contour, with pitch contours extracted using \textsc{SwiftF0}~\cite{nieradzik2025swiftf0}.

Although few participants reached this level of musical performance, the broader timeline (Fig.~\ref{fig:timeline}) shows that musical expression was not exclusive to experts. Even novices such as P1, P3, and P7 engaged in expressive moments, experimenting with recognizable pitch fragments, rhythmic phrasing, or playful melodic gestures.

\subsection{Final Design Strategies and Expressive Outcomes}
\label{subsec:final-design-strategies}

During the final design phase, participants adopted varied strategies shaped by the design prompt and their prior exploration. Four participants reused a previously assembled prototype---either as-is (P2, P4, P11) or with minor extension (P9)---sometimes framing it to fit the scenario despite partial alignment.

The remaining 8 built new instruments from scratch, with 6 explicitly drawing on earlier acoustic discoveries. P7 integrated a prior finding---submerging a resonator in water---to produce \textit{``strange and fun''} timbres, and P8 reinterpreted the \textit{bridge connector} as a tunable resonator based on earlier discoveries of its pitch-controlling behavior. 
In contrast, P6 and P10 emphasized imaginative shapes over acoustic refinement, aligning more with their audience goals than technical experimentation.

\begin{figure}[!htb]
\centering
  \includegraphics[width=1\columnwidth]{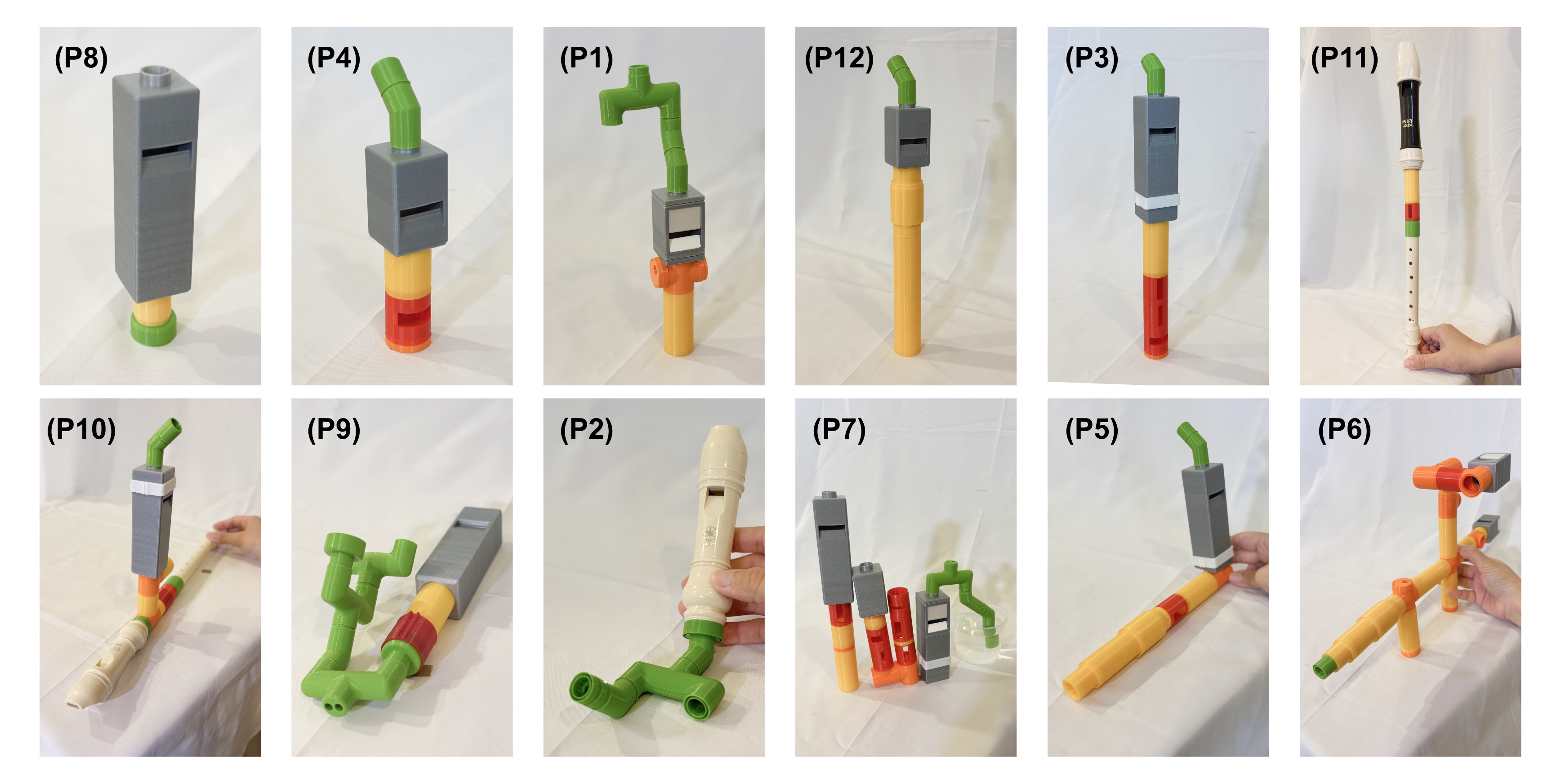} 
  \caption
  {%
    \textbf{Completed instruments from the scenario-based study.} All participants except P3, regardless of musical background, successfully built playable flutes using FlueBricks. P3 instead created a prank toy.
  }
  \Description
  {%
    Completed instruments from the scenario-based study. All participants except P3, regardless of musical background, successfully built playable flutes using FlueBricks. P3 instead created a prank toy.
  }
  \label{fig:result_works}
\end{figure}

These strategies yielded a wide range of expressive outcomes (Fig.~\ref{fig:result_works}). 
Four participants---P2, P8, P9, and P11---leveraged overtone behaviors to mimic animal sounds and explore timbral variation. 
Five others---P1, P4, P5, P8, and P12---focused on simplifying pitch control, either through minimal fingerings or telescoping body designs. 

P10 created a dual-generator flute for co-performance. P6 and P7 treated FlueBricks as sculptural media, constructing fantastical forms like a gun or castle to appeal to young learners. 
Only one participant, P3, failed to achieve meaningful acoustic control, interpreting the system as a prank toy. 
Still, 11 out of 12 participants produced musically or pedagogically intentional instruments, affirming FlueBricks’s capacity to support creative, goal-driven instrument construction.

\section{Discussion}
\label{sec:discussion}

Our findings revealed that tightly interleaving configuration and play naturally led participants to form and test hypotheses about acoustic behavior. We now discuss this emergent pattern---\textit{acoustic reasoning}---and examine the system's current limitations.

\subsection{Acoustic Reasoning}
\label{subsec:acoustic-reasoning}

As outlined in Section~\ref{sec:introduction}, we define \textit{acoustic reasoning} as an iterative process in which users form, test, and refine hypotheses about acoustic behavior through hands-on configuration and play. A complete \textit{acoustic reasoning cycle} comprises four stages---\textit{hypothesize, configure, evaluate, conclude} (Section~\ref{subsec:designer--player-loop}). This process requires an artifact that can be simultaneously configured and played, materializing acoustic parameters into reconfigurable parts. FlueBricks was designed around the \textit{designer--player loop}, tightly interleaving configuration and play; \textit{acoustic reasoning} emerged as the exploratory pattern participants naturally developed when these two activities became inseparable.

In practice, \textit{acoustic reasoning} is not an entirely new concept; it has long existed within the specialized domain of traditional instrument making. Boehm~\cite{Boehm2011} refined the flute by building detachable tube sections with movable tone holes, iteratively deriving acoustic principles---the \textit{rule formation} we observed (Section~\ref{subsec:rule-formation}). Similar iterative refinement appears in organ pipe voicing~\cite{Sakamoto2005, Rucz2014}. Conceptually, the roots of these practices lie in what Papert~\cite{Papert1980} termed ``objects-to-think-with''---the constructionist notion that learners deepen understanding by building and debugging external artifacts, much as Papert's gears made mathematical ratios tangible, FlueBricks makes acoustic principles configurable and playable. Overall, the contribution of FlueBricks is to transform such craft techniques---previously requiring expensive specialized craftsmanship---into accessible, modular forms, extending \textit{acoustic reasoning} from expert practice to general users' exploration.

Constructionism-based music research spans instrument performance learning~\cite{Ou2024}, musical concept learning~\cite{Downton2010}, and Digital Musical Instrument (DMI) design~\cite{Harriman2015, Ward2023, Jakobsen2016}, all sharing an emphasis on self-directed exploration through building and interacting with external artifacts. However, \textit{acoustic reasoning} differs from DMI design in that DMIs decouple interface from sound production, requiring learners to define arbitrary mappings~\cite{Harriman2015}; \textit{acoustic reasoning} instead grounds interaction in physical acoustics, letting users directly link structure to sound with immediate feedback and without complex circuitry~\cite{McPherson2015}. Choi and Kapur~\cite{Choi2022} identified that music students need active learning over instructional teaching; the \textit{acoustic reasoning cycle} responds through a dedicated build-and-play cycle. In acoustic education, LeMo~\cite{Arbel2022} similarly teaches through disassembling instrument components, but focuses on demonstrating acoustic behavior; our work instead examines how users actively construct knowledge through the iterative process of \textit{acoustic reasoning}.

To foster \textit{acoustic reasoning}, we deliberately excluded G2 paths of least resistance from Ledo et al.'s framework~\cite{Ledo18} (Section~\ref{sec:fluebricks}), while scaffolding through G1 easy to configure and play with, G4 familiar instrument anchors, and G5 flexible assembly. Excluding G2 strategically introduced what Gaver~\cite{Gaver2003} terms ``ambiguity as a design resource'': some modules lack direct counterparts in traditional instruments (Section~\ref{subsec:module-usage-patterns}), and removing prescriptive guidance encouraged users to enter the \textit{acoustic reasoning cycle} to understand module functions. Our empirical results (Section~\ref{subsec:reinterpretation}) confirmed that this design openness effectively triggered \textit{reinterpretation} behavior (Section~\ref{sec:fluebricks}).

To sustain \textit{acoustic reasoning}, design must balance physical fidelity and cognitive adaptivity. On the physical side, the artifact must eliminate unintended noise (Section~\ref{subsec:module-usage-patterns}) so users attribute errors to hypothesis revision rather than interface failure---echoing Papert's debuggability~\cite{Papert1980}. On the cognitive side, we propose \textit{adaptive ambiguity}: \textit{instrument scaffolding} provides cognitive anchors that accumulate into \textit{rule formation}, while physical ambiguity invites \textit{reinterpretation}. Our findings suggest that the effectiveness of this balance shifts with users' prior expertise (Section~\ref{subsec:perceptual-sensitivity}); designers should therefore calibrate ambiguity accordingly. Finally, \textit{performative acts} (Section~\ref{subsec:performative-act}) can be understood as consolidations of \textit{acoustic reasoning}---participants encoded fleeting acoustic discoveries into stable physical configurations, transforming them into shareable, iterable knowledge artifacts (Section~\ref{subsec:designer--player-loop}, Section~\ref{subsec:final-design-strategies}), echoing Papert's concept of ``public entities.'' Designs should therefore support mechanisms for capturing and sharing these configurations across users. Collectively, these three principles---physical fidelity, adaptive ambiguity, and performative consolidation---are particularly critical for acoustic tangible construction kits, where physical structure directly governs sound, auditory feedback is ephemeral, and the strong affordances of familiar instruments can both anchor learning and constrain exploration. We believe these principles may also be applicable to broader areas of sonic interaction and tangible computing.

\subsection{Limitations and Future Work}
\label{subsec:limitation-future-work}

A number of limitations shape FlueBricks's current scope. FFF 3D printing enabled rapid prototyping but introduced tolerance issues that can compromise the physical fidelity required for \textit{acoustic reasoning} (Section~\ref{subsec:acoustic-reasoning}). Future work will explore alternative materials and refined joint designs. The toolkit's range remains confined to flutes; future modules could extend into other aerophone mechanisms. Our evaluation involved 12 participants in single-session studies---sufficient to characterize exploratory behaviors but not broader pedagogical impact. Future studies may expand toward classroom settings with music educators, with optional scaffolding to support \textit{acoustic reasoning}.

Finally, our current design is physical-first. Future iterations may augment this foundation with computational hooks: logging configurations to visualize \textit{acoustic reasoning cycles}; tutorials that scaffold \textit{rule formation}; generating fabrication plans for stabilized designs; and user-guided interfaces for custom modules. These extensions would complement rather than replace the tangible interaction that enables \textit{acoustic reasoning}.

\section{Conclusion}
\label{sec:conclusion}

In conclusion, we introduced FlueBricks, a modular construction kit of \textit{generator}, \textit{resonator}, and \textit{connector} modules for \textit{acoustic reasoning} via building flute-like instruments within a \textit{designer--player loop}. In a 12-participant exploratory study, \textit{acoustic reasoning} manifested in \textit{instrument scaffolding}, \textit{rule formation}, \textit{reinterpretation}, and \textit{performative acts} across expertise levels. These findings suggest that balancing physical grounding with deliberate openness can foster \textit{acoustic reasoning}, informing future research on acoustic learning with constructive musical tools.

\begin{acks}
This work was supported by National Science and Technology Council in Taiwan (NSTC 113-2221-E-002 -185 -MY3). We thank our colleague Yu-Wei Chang for valuable insights during this research, and Zhu-Yuan Lai for helping us discuss and explore early prototypes.
\end{acks}

\balance
\bibliographystyle{ACM-Reference-Format}
\bibliography{reference}

\clearpage

\appendix
\section{Appendix}
\label{sec:appendix}

This appendix provides additional details on acoustic design rationale, module specifications, fabrication and assembly, usage configurations, and participant demographics.

\renewcommand\thetable{\thesection.\arabic{table}}
\setcounter{table}{0}

\subsection{Acoustic Design Rationale}
\label{app:acoustic-rationale}

\textbf{Generator.} The organ pipe, often described as the ``king of instruments,'' exemplifies how geometric modification of flute-like structures achieves wide tonal range~\cite{Douglas1965}. Beyond organology, traditions such as Native American flutes~\cite{Crawford2006} and recorder mouthpieces also highlight how small changes in fipple and labium geometry can significantly affect tonal character~\cite{AranaGarate2010}. These historical and cultural precedents informed our identification of the 6 internal geometries most critical to tone production, which guided the decomposition of the generator into 8 submodules.

\textbf{Resonator.} Musicians build on acoustic principles through fingerings, including cross-fingering~\cite{Adachi2016}, where combinations of open and closed holes allow bending notes, producing microtonal steps, and adjusting brightness. The shape of the resonator also influences timbre---globular forms such as the ocarina produce a softer quality than cylindrical pipes~\cite{Okada2019}. While FlueBricks primarily uses cylindrical segments, its voxel-based fabrication approach also permits constrained non-cylindrical forms, expanding the design space beyond traditional tubes while keeping modules structurally reliable. Beyond their individual roles, the resonator modules form a unified system for pitch resolution through a shared joint standard that allows parts to stack interchangeably, making the resonator a precise yet flexible space for constructing custom pitch systems.

\subsection{Module Specifications}
\label{app:module-specifications}

This section provides detailed specifications for all FlueBricks modules, organized by component family (Generator, Resonator, Connector) as introduced in Section~\ref{sec:fluebricks}.

\subsubsection{Generator Modules}

The generator family consists of 8 submodules that map to the six internal geometries most critical to tone production. Each submodule represents an acoustically meaningful unit, where swapping produces distinct, repeatable differences in sound.

\paragraph{\textbf{Air intake.}} 
Entry point for breath input, regulating airflow volume. Maps to the flue tunnel. Analogous to the ``foot'' in organ flue pipes~\cite{Steenbrugge2010}.

\paragraph{\textbf{Air duct support.}} 
Provides structural stability for airflow alignment. Not mapped to a single geometry, but maintains the integrity of the flue tunnel and ensures modular assembly.

\paragraph{\textbf{Air deflector.}} 
Redirects airflow toward the splitting edge, forming a buffer chamber. Maps to the air chamber, influencing onset reliability and tonal stability. Analogous to the languid in organ pipes~\cite{Mercer1951} and plugs in Native American flutes~\cite{Crawford2006}.

\paragraph{\textbf{Air facade.}} 
Forms the outer boundary of the windway, preserving jet direction. Maps to the windway, where outlet shaping affects tonal color and ease of speaking. Analogous to the lower lip in organ pipes~\cite{Nolle1979}.

\paragraph{\textbf{Window regulator.}} 
Adjusts the height and profile of the window, altering pitch and brightness. Maps directly to the window geometry~\cite{Nolle1979, Hruska2021}. Implemented in TPU to support real-time manual shaping and expressive exploration.

\paragraph{\textbf{Splitting edge.}} 
Divides the airstream to initiate oscillation. Maps to the window and labium region, influencing clarity, articulation, and harmonic content. Analogous to the upper lip in organ pipes~\cite{Nolle1979, Mercer1951}.

\paragraph{\textbf{Labium support.}} 
Stabilizes the splitting edge and shapes the cavity beneath the labium. Maps partially to the sound chamber. Provides alignment stability to ensure reliable tone production.

\paragraph{\textbf{Labium base.}} 
Forms the bottom of the generator and connects to the resonator. Maps to the sound chamber, anchoring the generator and setting the fundamental before further tuning~\cite{Halfpenny1956}.

\subsubsection{Resonator Modules}

The resonator family consists of 5 module types that enable users to control pitch and timbre through length and tone-hole manipulation. These modules support both discrete (\textit{basic nodes}) and continuous (\textit{telescope}) parameter exploration.

\paragraph{\textbf{Basic node.}} 
Foundational unit available in multiple diameters and lengths (e.g., \textit{M-short}, \textit{S-medium}). 
Defines the air column and sets the coarse pitch range. 
Extra-short variants allow hole placement near connection seams and enable fine-grained stacking resolution---up to 1/8 of a long node. 
This modular sizing echoes prior work on woodwind resonator design~\cite{Arbel2022, Allen2015}.

\paragraph{\textbf{Branch node.}} 
Adds a lateral tone hole that acts as a new termination point, directly shortening the effective length. 
By default, it provides a large circular hole and is designed to be paired with \textit{adapt caps}. 
This design was inspired by traditional woodwind side holes used to modulate pitch~\cite{Fletcher1998}.

\paragraph{\textbf{Adapt cap.}} 
Attaches to a \textit{branch node} to reduce hole size and add variability. 
Off-center openings allow the cap to be rotated, shifting hole placement around the tube wall. 
This pairing supports microtonal control and subtle timbral adjustments, a strategy reminiscent of tone-hole variation in flutes and recorders~\cite{Allen2015}.

\paragraph{\textbf{Tuning node.}} 
Replaces circular tone holes with vertical slots, inspired by tuning slots in organ flue pipes, traditionally used for pitch adjustment during voicing~\cite{Rucz2014}. 
Slot dimensions and orientations vary, supporting both fixed adjustments and expressive behaviors such as glissando. 
Designed to pair with the \textit{tweak ring} for live control.

\paragraph{\textbf{Tweak ring.}} 
A flexible TPU sleeve designed to pair with a \textit{tuning node}. 
Sliding or rotating it covers or reveals parts of the slot, turning a fixed opening into a continuously adjustable lever for live pitch control. 
TPU was chosen to ensure smooth, responsive interaction.

\subsubsection{Connector Modules}

The connector family consists of 3 modules that support airflow routing, branching structures, and hybrid configurations. These modules enable multi-voice instruments and integration with existing recorders.

\paragraph{\textbf{Air distribution hub.}} 
Central routing node that distributes airflow from one source across multiple outputs. 
Grounded in traditions such as drone flutes and bagpipes~\cite{Collinson1975}, it enables multi-chambered setups for dual-pitch instruments, layered articulations, or continuous drones.

\paragraph{\textbf{Air regulator.}} 
Modifies intake geometry to control airflow direction and comfort. 
Used standalone with a generator, it regulates the volume of incoming air and makes the instrument more ergonomic to blow from a human mouth (G1). 
In multi-voice configurations, pairing the regulator with the hub allows players to balance airflow across outputs for expressive performance (G5).

\paragraph{\textbf{Bridge connector.}} 
Ensures airtight connectivity across diverse geometries, linking FlueBricks modules or adapting them to standardized recorders (e.g., soprano, alto). 
Supports hybrid constructions such as recorder-FlueBricks extensions or dual-chamber flutes (G4).

\subsection{Fabrication and Assembly Details}
\label{app:fabrication}

All FlueBricks modules were fabricated using FFF/FDM 3D printing (Ultimaker S3/S5). We printed rigid parts in PLA and flexible parts (e.g., \textit{window regulator}, \textit{tweak ring}) in TPU. Our baseline print settings were a 0.1\,mm layer height and 20\% infill. Modules connect through press-fit joints: dual-wall fittings provide alignment and stability, while single-wall friction fits enable telescoping adjustments. We used color coding to indicate intended pairings (e.g., \textit{tuning node} + \textit{tweak ring}, \textit{branch node} + \textit{adapt cap}). For replication, we note that joint performance is sensitive to printer-dependent parameters such as nozzle/line width, wall count, support usage, and print orientation around the joint interface. We validated press-fit tolerances with small test prints and used light post-processing (e.g., sanding) when needed to restore smooth sliding and airtight seals.

\subsection{Additional Usage Configurations}
\label{app:design-configurations}

\begin{figure}[!htb]
\centering
  \includegraphics[width=1\columnwidth]{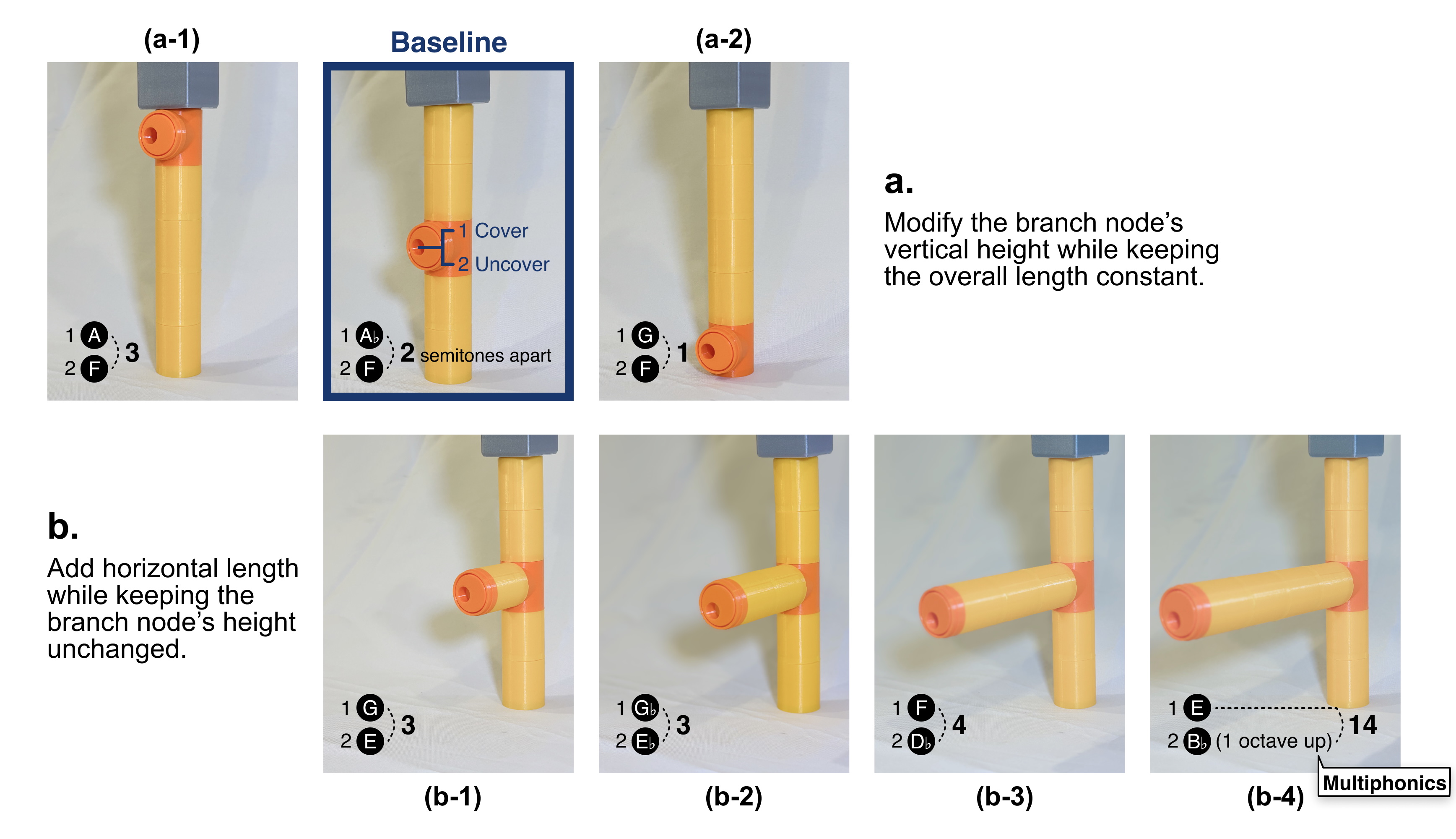}
  \caption{%
    \textbf{Pitch contrast and multiphonic effects through basic and branch node combinations.} (a) Raising the tone hole vertically (with fixed tube length) narrows the pitch gap between covered and uncovered states. (b) Extending the horizontal branch progressively lowers the covered tone (E4 → Eb4 → Db4 → Bb3) and eventually produces multiphonic effects when the upper note jumps an octave (G4 → F4 → F5), demonstrating how spatial layout enables musical expressivity.
  }
  \Description{%
    Pitch contrast and multiphonic effects through basic and branch node combinations. (a) Raising the tone hole vertically (with fixed tube length) narrows the pitch gap between covered and uncovered states. (b) Extending the horizontal branch progressively lowers the covered tone (E4 to E-flat 4 to D-flat 4 to B-flat 3) and eventually produces multiphonic effects when the upper note jumps an octave (G4 to F4 to F5), demonstrating how spatial layout enables musical expressivity.
  }
  \label{fig:walkthrough_3}
\end{figure}

Building on the tone-hole manipulation configuration from Section~\ref{sec:fluebricks-usage-configurations}, users can explore more advanced combinations by integrating the \textit{basic node} with the \textit{branch node} to shape pitch contrast and articulation behavior.
\\
They shift the tone hole vertically while keeping the total tube length fixed, discovering that a higher tone hole reduces the pitch gap between the covered and the uncovered states (Fig.~\ref{fig:walkthrough_3}a).
They keep the height constant but extend the side branch horizontally (Fig.~\ref{fig:walkthrough_3}b).
As the branch grows, the lower note drops progressively (\note[E4] → \note[Eb4] → \note[Db4] → \note[Bb3]), while the upper note descends gradually (\note[G4] → \note[Gb4] → \note[F4]) before suddenly jumping an octave to \note[F5], producing a distinct multiphonic effect (Fig.~\ref{fig:walkthrough_3}b-4).
\\
These combinations further tangibly prompt users to treat spatial layout as a tool for musical expressivity.

\begin{table*}[!b]
\subsection{Participant Demographics}
\label{app:participant-demographics}
\centering
\small
\setlength{\tabcolsep}{4pt}
\begin{tabular*}{\textwidth}{@{\extracolsep{\fill}}l c c p{4.2cm} p{1.8cm} c c p{1.6cm}@{}}
\toprule
\textbf{Participant} & \textbf{Age} & \textbf{Gender} & \textbf{Profession} & \textbf{Expertise Level} & \textbf{Creativity (Self)} & \textbf{Design Exp.} & \textbf{LEGO Play Style} \\
\midrule
P1 & 29 & Male & Engineer (IC design) & Novice & High & No & Exploratory \\
P2 & 36 & Female & Planner (marketing) & Amateur & Low & No & None \\
P3 & 30 & Male & Designer (educational games) & Novice & Low & No & Exploratory \\
P4 & 32 & Female & Teacher (early childhood) & Amateur & High & Yes & Exploratory \\
P5 & 51 & Female & Professor (innovation / design, flute) & Pro Woodwind & High & No & Goal-driven \\
P6 & 27 & Male & Student (music, dizi/suona) & Pro Woodwind & High & Yes & Exploratory \\
P7 & 32 & Female & Engineer (freelance) & Novice & Medium & Yes & Exploratory \\
P8 & 30 & Male & Consultant & Amateur & High & Yes & Goal-driven \\
P9 & 36 & Male & Artist (sound/new media) & Pro Musician & Low & Yes & Goal-driven \\
P10 & 26 & Male & Producer (music) & Pro Musician & Medium & Yes & Goal-driven \\
P11 & 34 & Female & Professor (music, composition) & Pro Musician & Low & Yes & None \\
P12 & 20 & Male & Student (music, recorder) & Pro Woodwind & High & Yes & Exploratory \\
\bottomrule
\end{tabular*}
\caption{Overview of participant demographics and creative backgrounds, including age, gender, profession, musical expertise, self-reported creativity, instrument design or modification experience, and LEGO play style, collected during the pre-interview. Instrument design/modification experience includes both casual and professional experiences. LEGO play style was used as a proxy for creative behavior, categorized as \textit{None}, \textit{Goal-driven} (structured builds), or \textit{Exploratory} (open-ended play).}
\Description{Overview of participant demographics and creative backgrounds for 12 participants (P1--P12), including age (20--51), gender, profession, musical expertise (Novice, Amateur, Pro Woodwind, or Pro Musician, three per group), self-reported creativity (Low, Medium, High), instrument design or modification experience (Yes or No), and LEGO play style (None, Goal-driven, or Exploratory), collected during the pre-interview. Instrument design or modification experience includes both casual and professional experiences. LEGO play style was used as a proxy for creative behavior. Participants span engineers, teachers, musicians, artists, and students.}
\label{tab:participant_background}
\end{table*}

\end{document}

%% file: _includes.tex
\usepackage{bm}
\usepackage{mathtools}
\usepackage{amsmath}
\usepackage{multirow}
\usepackage{xcolor}
\usepackage{tabularx}
\usepackage{balance}
\usepackage{booktabs}

\usepackage{multirow}
\usepackage{makecell}
\usepackage{stfloats}
\usepackage{soul}

\usepackage{graphicx}
\usepackage{color}
\usepackage{textcomp}
\usepackage{cuted}
\usepackage{enumerate}
\usepackage{enumitem}
\usepackage[all]{hypcap}    
\usepackage{subfigure}
\usepackage[utf8]{inputenc}
\usepackage{booktabs}

%% file: reference.bib
@article{turk1995oldest,
  title={The oldest musical instrument in Europe discovered in Slovenia},
  author={Turk, Ivan and Dirjec, Janez and Kavur, Boris},
  journal={Razprave IV. Razreda SAZU},
  volume={36},
  pages={287--293},
  year={1995}
}

@misc{ModularFluteNicolas2025,
	title = {Modular {Flute}},
	author = {{Nicolas} {Bras}},
	url = {https://www.kickstarter.com/projects/nicolasbras/modular-flute-by-nicolas-bras},
	language = {en},
	urldate = {2025-11-25},
	journal = {Kickstarter},
	month = oct,
	year = {2025},
}

@inproceedings{Hartmann2006,
  series = {UIST06},
  title = {Reflective physical prototyping through integrated design,  test,  and analysis},
  url = {http://dx.doi.org/10.1145/1166253.1166300},
  DOI = {10.1145/1166253.1166300},
  booktitle = {Proceedings of the 19th annual ACM symposium on User interface software and technology},
  publisher = {ACM},
  author = {Hartmann,  Bj\"{o}rn and Klemmer,  Scott R. and Bernstein,  Michael and Abdulla,  Leith and Burr,  Brandon and Robinson-Mosher,  Avi and Gee,  Jennifer},
  year = {2006},
  month = oct,
  collection = {UIST06}
}

@book{Papert1980,
  title = {Mindstorms: children, computers, and powerful ideas},
  publisher = {Basic Books},
  address = {New York, NY},
  author = {Papert, Seymour},
  year = {1980},
}

@book{Norman2013,
  author = {Norman, Donald A.},
  title = {The Design of Everyday Things},
  year = {2013},
  publisher = {Basic Books},
  address = {New York, NY},
  isbn = {978-0-465-05065-9}
}

@inproceedings{Dominiak2024,
  series = {DIS ’24},
  title = {ProtoBricks: A Research Toolkit for Tangible Prototyping \& Data Physicalization},
  url = {http://dx.doi.org/10.1145/3643834.3661573},
  DOI = {10.1145/3643834.3661573},
  booktitle = {Designing Interactive Systems Conference},
  publisher = {ACM},
  author = {Dominiak,  Julia and Walczak,  Anna and Stefanidi,  Evropi and Adamkiewicz,  Krzysztof and Grudzień,  Krzysztof and Niess,  Jasmin and Woźniak,  Paweł W.},
  year = {2024},
  month = jul,
  pages = {476–495},
  collection = {DIS ’24}
}

@inproceedings{Feick2023,
  series = {UIST ’23},
  title = {VoxelHap: A Toolkit for Constructing Proxies Providing Tactile and Kinesthetic Haptic Feedback in Virtual Reality},
  url = {http://dx.doi.org/10.1145/3586183.3606722},
  DOI = {10.1145/3586183.3606722},
  booktitle = {Proceedings of the 36th Annual ACM Symposium on User Interface Software and Technology},
  publisher = {ACM},
  author = {Feick,  Martin and Biyikli,  Cihan and Gani,  Kiran and Wittig,  Anton and Tang,  Anthony and Kr\"{u}ger,  Antonio},
  year = {2023},
  month = oct,
  pages = {1–13},
  collection = {UIST ’23}
}

@inproceedings{Lei2022,
  series = {CHI ’22},
  title = {O\&O: A DIY toolkit for designing and rapid prototyping olfactory interfaces},
  url = {http://dx.doi.org/10.1145/3491102.3502033},
  DOI = {10.1145/3491102.3502033},
  booktitle = {CHI Conference on Human Factors in Computing Systems},
  publisher = {ACM},
  author = {Lei,  Yuxuan and Lu,  Qi and Xu,  Yingqing},
  year = {2022},
  month = apr,
  pages = {1–21},
  collection = {CHI ’22}
}

@inproceedings{Greenberg2001,
  series = {UIST01},
  title = {Phidgets: easy development of physical interfaces through physical widgets},
  url = {http://dx.doi.org/10.1145/502348.502388},
  DOI = {10.1145/502348.502388},
  booktitle = {Proceedings of the 14th annual ACM symposium on User interface software and technology},
  publisher = {ACM},
  author = {Greenberg,  Saul and Fitchett,  Chester},
  year = {2001},
  month = nov,
  pages = {209–218},
  collection = {UIST01}
}

@article{Li2016,
  title = {Acoustic voxels: computational optimization of modular acoustic filters},
  volume = {35},
  ISSN = {1557-7368},
  url = {http://dx.doi.org/10.1145/2897824.2925960},
  DOI = {10.1145/2897824.2925960},
  number = {4},
  journal = {ACM Transactions on Graphics},
  publisher = {Association for Computing Machinery (ACM)},
  author = {Li,  Dingzeyu and Levin,  David I. W. and Matusik,  Wojciech and Zheng,  Changxi},
  year = {2016},
  month = jul,
  pages = {1–12}
}

@article{Montagu2017,
  title = {How Music and Instruments Began: A Brief Overview of the Origin and Entire Development of Music,  from Its Earliest Stages},
  volume = {2},
  ISSN = {2297-7775},
  url = {http://dx.doi.org/10.3389/fsoc.2017.00008},
  DOI = {10.3389/fsoc.2017.00008},
  journal = {Frontiers in Sociology},
  publisher = {Frontiers Media SA},
  author = {Montagu,  Jeremy},
  year = {2017},
  month = jun 
}

@article{Ward2023,
  title = {The development of a Modular Accessible Musical Instrument Technology Toolkit using action research},
  volume = {5},
  ISSN = {2624-9898},
  url = {http://dx.doi.org/10.3389/fcomp.2023.1113078},
  DOI = {10.3389/fcomp.2023.1113078},
  journal = {Frontiers in Computer Science},
  publisher = {Frontiers Media SA},
  author = {Ward,  Asha},
  year = {2023},
  month = oct 
}

@article{Kvifte2008,
  author    = {Tellef Kvifte},
  title     = {What is a musical instrument?},
  journal   = {Svensk Tidskrift för Musikforskning / Swedish Journal of Music Research},
  volume    = {90},
  pages     = {45--56},
  year      = {2008},
  url       = {https://publicera.kb.se/stm-sjm/article/view/34186}
}

@article{Tahrolu2020,
  title = {Digital Musical Instruments as Probes: How computation changes the mode-of-being of musical instruments},
  volume = {25},
  ISSN = {1469-8153},
  url = {http://dx.doi.org/10.1017/S1355771819000475},
  DOI = {10.1017/s1355771819000475},
  number = {1},
  journal = {Organised Sound},
  publisher = {Cambridge University Press (CUP)},
  author = {Tahıroğlu,  Koray and Magnusson,  Thor and Parkinson,  Adam and Garrelfs,  Iris and Tanaka,  Atau},
  year = {2020},
  month = mar,
  pages = {64–74}
}

@article{Rogers1977,
  title = {Instrument Craftsman},
  volume = {63},
  ISSN = {1945-0087},
  url = {http://dx.doi.org/10.2307/3395233},
  DOI = {10.2307/3395233},
  number = {7},
  journal = {Music Educators Journal},
  publisher = {SAGE Publications},
  author = {Rogers,  Charlie},
  year = {1977},
  month = mar,
  pages = {120–123}
}

@misc{OpenFabPDX2025,
  author = {{OpenFab PDX}},
  title = {Modular Fiddle: Open Source Digital Modular Violin},
  year = {2025},
  url = {https://openfabpdx.com/modular-fiddle/},
  note = {Accessed: March 31, 2025},
}

@article{Arbel2022,
  title = {LeMo: an assembly kit for musical acoustics education},
  volume = {51},
  ISSN = {1744-5027},
  url = {http://dx.doi.org/10.1080/09298215.2022.2150651},
  DOI = {10.1080/09298215.2022.2150651},
  number = {2–3},
  journal = {Journal of New Music Research},
  publisher = {Informa UK Limited},
  author = {Arbel,  Lior and Gautier,  Fran\c{c}ois},
  year = {2022},
  month = may,
  pages = {106–120}
}

@article{Terrien2013,
  title = {Flute-like musical instruments: A toy model investigated through numerical continuation},
  volume = {332},
  ISSN = {0022-460X},
  url = {http://dx.doi.org/10.1016/j.jsv.2013.01.041},
  DOI = {10.1016/j.jsv.2013.01.041},
  number = {15},
  journal = {Journal of Sound and Vibration},
  publisher = {Elsevier BV},
  author = {Terrien,  Soizic and Vergez,  Christophe and Fabre,  Benoît},
  year = {2013},
  month = jul,
  pages = {3833–3848}
}

@article{Pal2006,
  title = {A combined LDA and flow-visualization study on flue organ pipes},
  volume = {40},
  ISSN = {1432-1114},
  url = {http://dx.doi.org/10.1007/s00348-006-0114-0},
  DOI = {10.1007/s00348-006-0114-0},
  number = {6},
  journal = {Experiments in Fluids},
  publisher = {Springer Science and Business Media LLC},
  author = {Paál,  G. and Angster,  J. and Garen,  W. and Miklós,  A.},
  year = {2006},
  month = may,
  pages = {825–835}
}

@article{Zoran2011,
  title = {The 3D Printed Flute: Digital Fabrication and Design of Musical Instruments},
  volume = {40},
  ISSN = {1744-5027},
  url = {http://dx.doi.org/10.1080/09298215.2011.621541},
  DOI = {10.1080/09298215.2011.621541},
  number = {4},
  journal = {Journal of New Music Research},
  publisher = {Informa UK Limited},
  author = {Zoran,  Amit},
  year = {2011},
  month = dec,
  pages = {379–387}
}

@article{Kolomiets2020,
  title = {The titanium 3D-printed flute: New prospects of additive manufacturing for musical wind instruments design},
  volume = {50},
  ISSN = {1744-5027},
  url = {http://dx.doi.org/10.1080/09298215.2020.1824240},
  DOI = {10.1080/09298215.2020.1824240},
  number = {1},
  journal = {Journal of New Music Research},
  publisher = {Informa UK Limited},
  author = {Kolomiets,  A. and Grobman,  Y.J. and Popov,  V. V. and Strokin,  E. and Senchikhin,  G. and Tarazi,  E.},
  year = {2020},
  month = sep,
  pages = {1–17}
}

@article{Umetani2016,
  title = {Printone: interactive resonance simulation for free-form print-wind instrument design},
  volume = {35},
  ISSN = {1557-7368},
  url = {http://dx.doi.org/10.1145/2980179.2980250},
  DOI = {10.1145/2980179.2980250},
  number = {6},
  journal = {ACM Transactions on Graphics},
  publisher = {Association for Computing Machinery (ACM)},
  author = {Umetani,  Nobuyuki and Panotopoulou,  Athina and Schmidt,  Ryan and Whiting,  Emily},
  year = {2016},
  month = nov,
  pages = {1–14}
}

@book{Fletcher1998,
  title = {The Physics of Musical Instruments},
  ISBN = {9780387216034},
  url = {http://dx.doi.org/10.1007/978-0-387-21603-4},
  DOI = {10.1007/978-0-387-21603-4},
  publisher = {Springer New York},
  author = {Fletcher,  Neville H. and Rossing,  Thomas D.},
  year = {1998}
}

@article{Nesterova_2020,
    author = {Nesterova, Olesya and Shapilov, Vitaliy and Begembetova, Galiya and Abdirakhman, Gulnar and Abdrashev, Bakhtiyar and Glebov, Victor},
    title = {The main trends in the modern techniques of playing the wind instrument},
    journal = {Opción: Revista de Ciencias Humanas y Sociales},
    volume = {91},
    year = {2020},
    pages = {405--424}
}

@book{Douglas1965,
  author    = {Douglas, George Ashdown},
  title     = {The Art of Organ-building: A Comprehensive Historical, Theoretical, and Practical Treatise on the Tonal Appointment and Mechanical Construction of Concert-room, Church, and Chamber Organs},
  publisher = {Dover Publications},
  year      = {1965},
  address   = {New York, NY, USA},
  volume    = {2},
  pages     = {758}
}

@book{Crawford2006,
  author       = {Tim R. Crawford and Kathleen Joyce-Grendahl},
  title        = {Flute Magic: An Introduction to the Native American Flute},
  edition      = {3rd},
  year         = {2006},
  publisher    = {Mel Bay Publications, Inc.},
  address      = {Pacific, MO},
}

@inproceedings{Steenbrugge2010,
  author = {Steenbrugge, D.},
  title = {Flow Acoustical Determinants of Historic Flue Organ Pipe Voicing Practices},
  booktitle = {Proceedings of the Associated Meeting of the International Congress on Acoustics (ICA)},
  year = {2010},
  address = {Sydney and Katoomba, Australia},
  month = {August 25--31},
  organization = {International Congress on Acoustics},
}

@article{Mercer1951,
  title = {The Voicing of Organ Flue Pipes},
  volume = {23},
  ISSN = {1520-8524},
  url = {http://dx.doi.org/10.1121/1.1906727},
  DOI = {10.1121/1.1906727},
  number = {1},
  journal = {The Journal of the Acoustical Society of America},
  publisher = {Acoustical Society of America (ASA)},
  author = {Mercer,  Derwent M. A.},
  year = {1951},
  month = jan,
  pages = {45–54}
}

@article{Nolle1979,
  title = {Some voicing adjustments of flue organ pipes},
  volume = {66},
  ISSN = {1520-8524},
  url = {http://dx.doi.org/10.1121/1.383658},
  DOI = {10.1121/1.383658},
  number = {6},
  journal = {The Journal of the Acoustical Society of America},
  publisher = {Acoustical Society of America (ASA)},
  author = {Nolle,  A. W.},
  year = {1979},
  month = dec,
  pages = {1612–1626}
}

@inproceedings{Hruska2021,
  author = {Hruška, V. and Dlask, P.},
  title = {Organ Pipe Voicing and Heuristic Optimization},
  booktitle = {Proceedings of the Annual Congress of the International Institute of Acoustics and Vibration (IIAV)},
  year = {2021},
  location = {Prague, Czech Republic},
  publisher = {International Institute of Acoustics and Vibration},
}

@article{Halfpenny1956,
  title = {The English Baroque Treble Recorder},
  volume = {9},
  ISSN = {0072-0127},
  url = {http://dx.doi.org/10.2307/841791},
  DOI = {10.2307/841791},
  journal = {The Galpin Society Journal},
  publisher = {JSTOR},
  author = {Halfpenny,  Eric},
  year = {1956},
  month = jun,
  pages = {82}
}

@article{Allen2015,
  title = {Aerophones in flatland: interactive wave simulation of wind instruments},
  volume = {34},
  ISSN = {1557-7368},
  url = {http://dx.doi.org/10.1145/2767001},
  DOI = {10.1145/2767001},
  number = {4},
  journal = {ACM Transactions on Graphics},
  publisher = {Association for Computing Machinery (ACM)},
  author = {Allen,  Andrew and Raghuvanshi,  Nikunj},
  year = {2015},
  month = jul,
  pages = {1–11}
}

@inproceedings{AranaGarate2010,
  author    = {Juan Luis Arana-Garate and Juan Angel Asensio},
  title     = {Effect of jet pressure in musical instruments of the txistu family},
  booktitle = {Proceedings of INTER-NOISE and NOISE-CON Congress and Conference},
  year      = {2010},
  number    = {5},
  pages     = {4788--4795},
}

@article{Rucz2014,
  title = {Acoustic behavior of tuning slots of labial organ pipes},
  volume = {135},
  ISSN = {1520-8524},
  url = {http://dx.doi.org/10.1121/1.4869679},
  DOI = {10.1121/1.4869679},
  number = {5},
  journal = {The Journal of the Acoustical Society of America},
  publisher = {Acoustical Society of America (ASA)},
  author = {Rucz,  Péter and Augusztinovicz,  F\"{u}l\"{o}p and Angster,  Judit and Preukschat,  Tim and Miklós,  András},
  year = {2014},
  month = may,
  pages = {3056–3065}
}

@book{Collinson1975,
  author    = {Francis M. Collinson},
  title     = {The Bagpipe: The History of a Musical Instrument},
  publisher = {Routledge \& K. Paul},
  year      = {1975},
  address   = {London, UK}
}

@article{Southcott2016Early,
  author = {Jane Southcott},
  title = {Early Days of Recorder Teaching in South Australian Schools: A Personal History},
  journal = {Australian Journal of Music Education},
  year = {2016},
  volume = {50},
  pages = {16-26},
  url = {https://www.semanticscholar.org/paper/Early-Days-of-Recorder-Teaching-in-South-Australian-Southcott/6192d2291bfb16037f8835de1dcc004caa5c402a}
}

@misc{YamahaCorporationndSchool,
  author = {Yamaha Corporation},
  title = {The School Project},
  url = {https://www.yamaha.com/en/stories/culture/school-project/},
  year = {n.d.}
}

@misc{YamahaCorporationndCreating,
  author = {Yamaha Corporation},
  title = {Creating Bonds by Expanding Musical Opportunities - The Key},
  url = {https://www.yamaha.com/en/stories/the-key/006-01/},
  year = {n.d.}
}

@inproceedings{Ledo18,
author = {Ledo, David and Houben, Steven and Vermeulen, Jo and Marquardt, Nicolai and Oehlberg, Lora and Greenberg, Saul},
title = {Evaluation Strategies for HCI Toolkit Research},
year = {2018},
isbn = {9781450356206},
publisher = {Association for Computing Machinery},
address = {New York, NY, USA},
url = {https://doi.org/10.1145/3173574.3173610},
doi = {10.1145/3173574.3173610},
booktitle = {Proceedings of the 2018 CHI Conference on Human Factors in Computing Systems},
pages = {1–17},
numpages = {17},
location = {Montreal QC, Canada},
series = {CHI '18}
}

@article{Adachi2016,
  title = {A simple model of cross-fingering explaining both pitch flattening and sharpening},
  volume = {140},
  ISSN = {1520-8524},
  url = {http://dx.doi.org/10.1121/1.4971029},
  DOI = {10.1121/1.4971029},
  journal = {Journal of the Acoustical Society of America},
  publisher = {Acoustical Society of America (ASA)},
  author = {Adachi, Seiji},
  year = {2016},
  month = {oct},
  pages = {3426--3427}
}

@inproceedings{Okada2019,
  author    = {Hiroaki Okada and Sho Iwagami and Taizo Kobayashi and Kinya Takahashi},
  title     = {Numerical Simulation of Aerodynamics Sound in an Ocarina Model},
  booktitle = {Proceedings of the International Symposium on Musical Acoustics (ISMA 2019)},
  year      = {2019},
  pages     = {263--268},
  address   = {Detmold, Germany}
}

@misc{nieradzik2025swiftf0,
      title={SwiftF0: Fast and Accurate Monophonic Pitch Detection},
      author={Lars Nieradzik},
      year={2025},
      eprint={2508.18440},
      archivePrefix={arXiv},
      primaryClass={cs.SD},
      url={https://arxiv.org/abs/2508.18440},
}

@book{Wiberg2018,
  title = {The Materiality of Interaction: Notes on the Materials of Interaction Design},
  ISBN = {9780262344692},
  url = {http://dx.doi.org/10.7551/mitpress/9780262037518.001.0001},
  DOI = {10.7551/mitpress/9780262037518.001.0001},
  publisher = {The MIT Press},
  author = {Wiberg,  Mikael},
  year = {2018},
  month = feb 
}

@article{Lindell2013,
  title = {Crafting interaction: The epistemology of modern programming},
  volume = {18},
  ISSN = {1617-4917},
  url = {http://dx.doi.org/10.1007/s00779-013-0687-6},
  DOI = {10.1007/s00779-013-0687-6},
  number = {3},
  journal = {Personal and Ubiquitous Computing},
  publisher = {Springer Science and Business Media LLC},
  author = {Lindell,  Rikard},
  year = {2013},
  month = may,
  pages = {613–624}
}

@article{McCarthy2004,
  title = {Technology as experience},
  volume = {11},
  ISSN = {1558-3449},
  url = {http://dx.doi.org/10.1145/1015530.1015549},
  DOI = {10.1145/1015530.1015549},
  number = {5},
  journal = {Interactions},
  publisher = {Association for Computing Machinery (ACM)},
  author = {McCarthy,  John and Wright,  Peter},
  year = {2004},
  month = sep,
  pages = {42–43}
}

@inbook{OModhrain2018,
  title = {Once More,  with Feeling: Revisiting the Role of Touch in Performer-Instrument Interaction},
  ISBN = {9783319583167},
  ISSN = {2192-2985},
  url = {http://dx.doi.org/10.1007/978-3-319-58316-7_2},
  DOI = {10.1007/978-3-319-58316-7_2},
  booktitle = {Musical Haptics},
  publisher = {Springer International Publishing},
  author = {O’Modhrain,  Sile and Gillespie,  R. Brent},
  year = {2018},
  pages = {11–27}
}

@article{Leonard2020,
  title = {Multisensory instrumental dynamics as an emergent paradigm for digital musical creation: A retrospective and prospective of haptic-audio creation with physical models},
  volume = {14},
  ISSN = {1783-8738},
  url = {http://dx.doi.org/10.1007/s12193-020-00334-y},
  DOI = {10.1007/s12193-020-00334-y},
  number = {3},
  journal = {Journal on Multimodal User Interfaces},
  publisher = {Springer Science and Business Media LLC},
  author = {Leonard,  James and Villeneuve,  Jér\^ome and Kontogeorgakopoulos,  Alexandros},
  year = {2020},
  month = jul,
  pages = {235–253}
}

@inproceedings{Gaver2003,
  series = {CHI03},
  title = {Ambiguity as a resource for design},
  url = {http://dx.doi.org/10.1145/642611.642653},
  DOI = {10.1145/642611.642653},
  booktitle = {Proceedings of the SIGCHI Conference on Human Factors in Computing Systems},
  publisher = {ACM},
  author = {Gaver,  William W. and Beaver,  Jacob and Benford,  Steve},
  year = {2003},
  month = apr,
  pages = {233–240},
  collection = {CHI03}
}

@article{Choi2022,
  title = {Human-centered design in acoustics education for undergraduate music majors},
  volume = {151},
  ISSN = {1520-8524},
  url = {http://dx.doi.org/10.1121/10.0010043},
  DOI = {10.1121/10.0010043},
  number = {4},
  journal = {The Journal of the Acoustical Society of America},
  publisher = {Acoustical Society of America (ASA)},
  author = {Choi,  Minsik and Kapur,  Max},
  year = {2022},
  month = apr,
  pages = {2282–2289}
}

@phdthesis{Harriman2015,
  author       = {Harriman, Jr., Jeffrey Wood},
  title        = {The Development and Use of Scaffolded Design Tools for Interactive Music},
  school       = {University of Colorado Boulder},
  year         = {2015},
  address      = {Boulder, CO, USA},
  type         = {{Ph.D.} Dissertation},
  url          = {https://www.colorado.edu/atlas/sites/default/files/attached-files/jifferharriman-dissertation-small_0.pdf}
}

@article{Ou2024,
  author = {Ou, Jiayi},
  title = {Practice design of violin teaching model based on constructivism learning theory},
  journal = {International Journal of New Developments in Education},
  volume = {6},
  number = {9},
  pages = {224--228},
  year = {2024},
  doi = {10.25236/IJNDE.2024.060934},
  publisher = {Francis Academic Press},
  ISSN = {2663-8169}
}

@techreport{Downton2010,
  author = {Downton, Michael P. and Peppler, Kylie A. and Portowitz, Adena},
  title = {Building Tunes Block by Block: Constructing Musical and Cross-Cultural Understanding through Impromptu},
  year = {2010},
  institution = {Indiana University},
  url = {https://kyliepeppler.com/Docs/2010_Peppler_Building_Tunes.pdf}
}

@inproceedings{Sakamoto2005,
  author = {Sakamoto, Yumiko and Yoshikawa, Shigeru and Angster, Judit},
  title = {Acoustical Investigations on the Ears of Flue Organ Pipes},
  booktitle = {Forum Acusticum},
  year = {2005},
  url = {https://www.academia.edu/18627118/Acoustical_investigations_on_the_ears_of_flue_organ_pipes}
}

@book{Boehm2011,
  author = {Boehm, Theobald},
  title = {The Flute and Flute Playing},
  publisher = {Courier Corporation},
  year = {2011},
  isbn = {9780486212857},
  url = {https://ia801308.us.archive.org/22/items/cu31924021743822/cu31924021743822.pdf}
}

@inproceedings{McPherson2015,
  author = {McPherson, Andrew and Zappi, Victor},
  title = {Exposing the Scaffolding of Digital Instruments with Hardware-Software Feedback Loops},
  booktitle = {Proceedings of the International Conference on New Interfaces for Musical Expression},
  year = {2015},
  pages = {258--263},
  url = {https://www.nime.org/proceedings/2015/nime2015_258.pdf},
  doi = {10.5281/zenodo.1179194}
}

@inproceedings{Jakobsen2016,
  author = {Kasper Buhl Jakobsen and Marianne Graves Petersen and Majken Kirkegaard Rasmussen and Jens Emil Gr{\o}nbaek and Jakob Winge and Jeppe Stougaard},
  title = {Hitmachine: Collective Musical Expressivity for Novices},
  booktitle = {Proceedings of the International Conference on New Interfaces for Musical Expression},
  pages = {241--246},
  year = {2016},
  doi = {10.5281/zenodo.1176038},
  url = {https://zenodo.org/record/1176038},
  address = {Brisbane, Australia}
}
